\documentclass[aps,pre,preprint,superscriptaddress]{revtex4-2}
\usepackage{graphicx,color}
\usepackage{amssymb,amsmath}

\begin{document}


\title{Particle trajectories, gamma-ray emission, and anomalous radiative trapping effects in magnetic dipole wave}


\author{A.~V.~Bashinov}
\email{bashinov@ipfran.ru}
\author{E.~S.~Efimenko}
\author{A.~A.~Muraviev }
\affiliation{Institute of Applied Physics, Russian Academy of Sciences, Nizhny Novgorod 603950, Russia}

\author{V.~D.~Volokitin}
\author{I.~B.~Meyerov}
\affiliation{Lobachevsky State University of Nizhny Novgorod, Nizhny Novgorod 603950, Russia}

\author{G.~Leuchs}
\affiliation{Institute of Applied Physics, Russian Academy of Sciences, Nizhny Novgorod 603950, Russia}
\affiliation{Max Planck Institute for the Science of Light, Erlangen, Germany}

\author{A.~M.~Sergeev}
\author{A.~V.~Kim}
\affiliation{Institute of Applied Physics, Russian Academy of Sciences, Nizhny Novgorod 603950, Russia}


\date{\today}

\begin{abstract}
In studies of interaction of matter with laser fields of extreme intensity there are two limiting cases of a multi-beam setup maximizing either the electric field or the magnetic field. In this work attention is paid to the optimal configuration of laser beams in the form of an m-dipole wave, which maximizes the magnetic field. We consider in such highly inhomogeneous fields the advantages and specific features of laser-matter interaction, which stem from individual particle trajectories that are strongly affected by gamma photon emission. It is shown that in this field mode qualitatively different scenarios of particle dynamics take place in comparison with the mode that maximizes the electric field. A detailed map of possible regimes of particle motion ({\it ponderomotive trapping}, {\it normal radiative trapping}, {\it radial} and {\it axial anomalous radiative trapping}), as well as angular and energy distributions of particles and gamma photons, is obtained in a wide range of laser powers up to 300~PW and reveals signatures of radiation losses experimentally detectable even with subpetawatt lasers.
\end{abstract}


\maketitle

\section{Introduction\label{Sec1}}

Currently, actively developing multipetawatt laser facilities \cite{Danson_HPLSE2019} can become a unique tool for studying the properties of quantum vacuum and quantum electrodynamics (QED) processes in extremely strong fields \cite{Marklund_RMP2006,Piazza_RMP2012}, as well as for modeling astrophysical phenomena in laboratory conditions  \cite{Chen_EPJST2014,Liang_SR2015,Takabe_HPLSE2021}. Previously, the main attention has been paid to optimal configurations of the multibeam laser setup in which the electric field attains the maximal value. In this case the electric field assumes the leading role in establishing the main physical effects, for example, the direct tunneling ionization of vacuum by laser pulses \cite{Narozhny_JETPL2004,Fedotov_LP2009,Bulanov_PRL2010M,Bulanov_PRL2010S,Gonoskov_PRL2013}. For such field structures detailed studies have also been done with respect to gamma-ray emission, QED cascades and electron-positron pair plasma production (see also \cite{Bell_PRL2008,Fedotov_PRL2010,Nerush_PRL2011,Bashinov_POP2013,Gelfer_PRA2015,Luo_POP2015,Jirka_SR2017,Vranic_PPCF2017,Tamburini_SR2017,Gong_PRE2017,Gonoskov_PRX2017}). Particularly, in the limiting case of a converging e-dipole wave \cite{Gonoskov_PRA2012}, characterized by a minimal focal volume and the highest electric field, extreme plasma states can be created, paving the way to the quantum pair plasma and the Schwinger field \cite{Efimenko_SR2018,Efimenko_PRE2019}. However, there are also many fundamental effects in which the magnetic field is a key factor, for example, high energy electrons radiate gamma photons mainly on the curved part of their trajectories induced by a strong magnetic field \cite{Landau2}. Consequently, the choice of the field structure in a laser multibeam configuration should depend on the specific physical effects in extreme laser-matter interactions we are interested in.

In this paper we pay attention to the case of tightly focused laser beams that maximize the magnetic field, and begin our consideration with the limiting case in the form of a converging m-dipole wave. We keep in mind such an important problem as vacuum breakdown in tightly focused laser beams and dense pair plasma production in the laboratory \cite{Bulanov_PRL2010M,Nerush_PRL2011,Gelfer_PRA2015,Vranic_PPCF2017,Efimenko_SR2018,Efimenko_PRE2019,Zhu_NC2016,Yuan_PPCF2018,Ridgers_PRL2012,Lecz_PPCF2019,Mironov_PRA2021}. In this problem both electric and magnetic fields are equally important. An electron is accelerated to high energy in the driving electric field, but basically it emits a gamma photon on a curved trajectory in the strong magnetic field \cite{Bashmakov_POP2014,Gonoskov_PRL2014}. The gamma photon mainly decays in strong magnetic field into a pair of an electron and a positron \cite{Bashmakov_POP2014}, which, in turn, may get accelerated to high energies by the electric field. Thus, both electric and magnetic domains in space and time contribute significantly  \cite{Bashmakov_POP2014} to the pair avalanche \cite{Bell_PRL2008} in tightly focused beams. Our aim is to show advantages and specific features of the laser-matter interaction at extreme intensities in highly inhomogeneous fields of the m-dipole wave, identify the experimental signatures for detecting these features, as well as investigate new pair plasma states that can be produced in the laboratory. The problem is complex and, since laser-matter interaction in these fields has not yet been investigated, certain studies should be completed from the ground up, starting from single particle trajectories and gamma photon emission, following with vacuum breakdown and the nonlinear stage of laser plasma interactions. This paper is devoted to studying particle trajectories in such highly inhomogeneous fields in wide range of powers from 0.01~PW to 300~PW. Trajectories not only shape the evolution of particle ensemble but also define QED processes: particles emit gamma photons which in turn can decay into electron-positron pairs in laser fields \cite{Nikishov_1967}.

This paper is organized as follows. In Sec.~\ref{Sec2_1}, we briefly discuss the standing wave approach in the analysis of particle motion and pair plasma generation, and in Sec.~\ref{Sec2_2} we describe the distributions of electric and magnetic fields of the m-dipole wave. In Sec.~\ref{Sec3_1} we present general properties of particle motion. In Sec.~\ref{Sec3_2} and \ref{Sec3_3} based on the time evolution of the particle ensemble, its spatial distribution and energy and angular spectra we propose several characteristics in order to reveal distinctions in particle motion caused by radiation losses and determine threshold powers of possible modes of motion. In Sec.~\ref{Sec3_4} we analyze in more detail particle trajectories within the determined characteristic power ranges. In Sec.~\ref{Sec3_5} properties of the gamma emission generated on these trajectories are considered. In Sec.~\ref{Sec4}, based on specific features of energy and angular distributions of particles and photons, an experimental approach for detection of signatures of radiation losses is proposed. Finally, in Sec.~\ref{Sec5} the results are summarized.

\section{Formulation of the problem\label{Sec2}}

\subsection{General approach\label{Sec2_1}}

In order to generate pair plasma by incoming laser pulses we must prepare a seed in the focal region to trigger vacuum breakdown. Here the seed can be an electron beam, a plasma or solid target, or a gamma ray (see for an example \cite{Sokolov_PRL2010,Mironov_PLA2014,Samsonov_SR2019,Artemenko_PRA2017,Slade_NJP2019,Mironov_PRA2013}). In general, this problem looks like the well-known air breakdown in a focused laser beam. However, at relativistically strong intensities seed parameters or even its electrodynamic properties can be greatly changed under the influence of incident fields and this can affect subsequent dynamics of interaction. For example, seed electrons can be accelerated to high energies and pushed to the focal point by ponderomotive forces of the incident laser pulses \cite{Gonoskov_PRL2014,Bashinov_QE2013} that may influence the generated gamma photons as well as pair production. In this case, the interaction scenario depends on many parameters of the seed and the laser, such as the parameters of the particle distribution, the shape of the pulse or even the turn-on duration etc., which makes the general analysis uninformative, depriving us of a qualitative understanding of possible development scenarios. However, in the case of interest, i.e., in tight focusing geometry, there are factors that can greatly simplify the formulation of the problem. A common factor for any type of mode is that the actual interaction volume is very small, much less than $\lambda^3$, where $\lambda$ is the laser wavelength. In this case we can consider the trajectories of particles in the standing wave. Excluding the nonlinear stage of interaction, we can make a general analysis of possible regimes of motion depending on the laser power in given fields of the standing wave. Using PIC simulations, we tested this idea for incoming laser pulses and confirmed that the standing wave approach works well.

\subsection{Field structure of the m-dipole wave\label{Sec2_2}}

For clarity we briefly introduce the field structure of the standing m-dipole wave and describe its features. Its magnetic field has a poloidal structure and the electric field is toroidal. The field structure possesses axial symmetry. We define a point on the axis of symmetry, where the magnetic field amplitude is maximal, as the central point and let it coincide with the coordinate origin. Without loss of generality let us assume that the axis of symmetry is the $z$ axis and the magnetic field in the central point is also directed along it. Both electric and magnetic fields are strongest at the central plane passing through the central point perpendicular to the $z$ axis. The main component of the magnetic field in the vicinity of the central point is the $z$ component. Outside the central plane due to the poloidal structure the magnetic field also has a radial component in the cylindrical coordinate system. The electric field is purely azimuthal.

\begin{figure}[t]
	\includegraphics[width = 1\columnwidth]{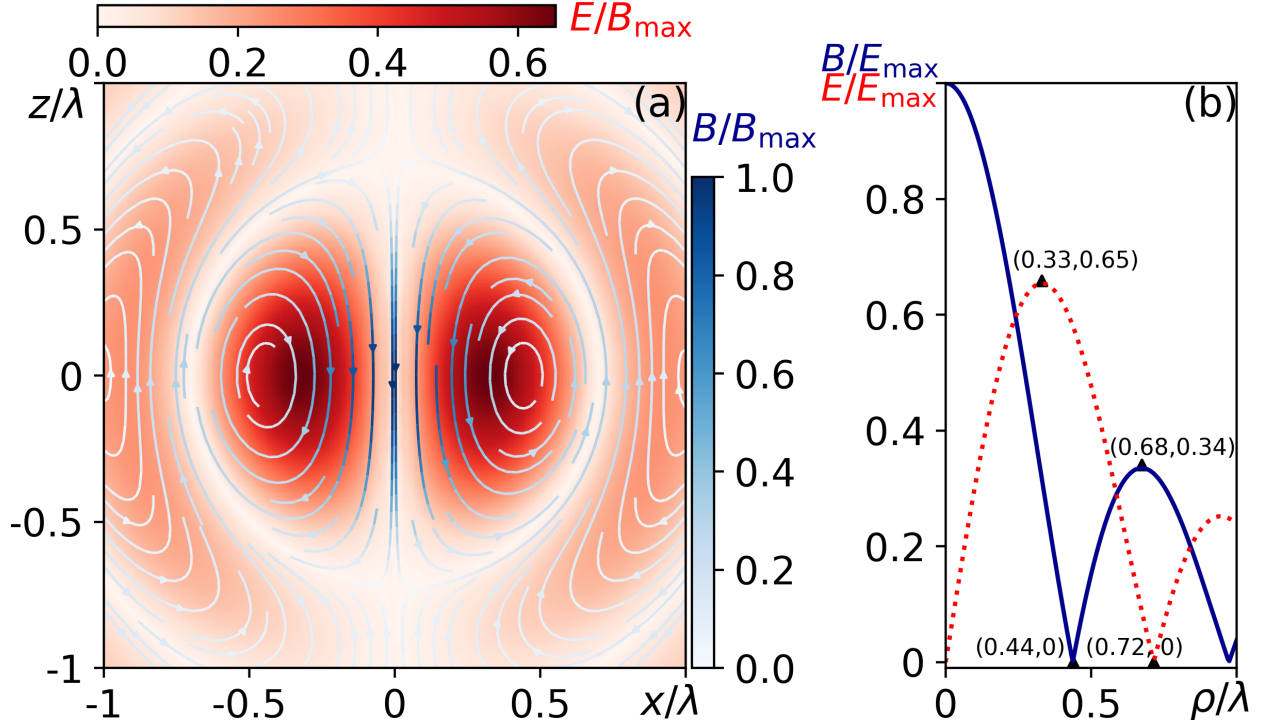}
	\caption{(Color online) Spatial distribution of electric and magnetic fields of the standing m-dipole wave. The electric field distribution and magnetic field lines are presented in the figure (a) in coordinates $z$ and $x$ in the Cartesian coordinate system. Electric (dotted line) and magnetic (solid line) fields as function of $\rho$ in the central plane are shown in the figure (b). Triangle markers denote locations and values of field extrema.\label{FigFields}}
\end{figure}

The exact analytical expressions for fields of the standing m-dipole wave are following \cite{Gonoskov_PRA2012}:
\begin{equation}
	\begin{aligned}
		\mathbf{E}=&F_0\sqrt{P_{\mathrm{PW}}}\cos{(\omega t)}\frac{\rho}{R}\left[\frac{\sin{\left(kR\right)}}{\left(kR\right)^2}-\frac{\cos{\left(kR\right)}}{kR}\right]\mathbf{e_\varphi};\\
		\mathbf{B}=&-F_0\sqrt{P_{\mathrm{PW}}}\sin{(\omega t)}\left\{\left[\frac{\sin{\left(kR\right)}}{kR}\left(1-\frac{1+(kz)^2}{\left(kR\right)^2}+\frac{3(kz)^2}{\left(kR\right)^4}\right)+\right.\right.\\
		&\left.\left.\frac{\cos{\left(kR\right)}}{\left(kR\right)^2}\left(1-\frac{3(kz)^2}{\left(kR\right)^4}\right)\right]\mathbf{e_z}+\frac{\rho z}{R^2}\left[\frac{\sin{\left(kR\right)}}{kR}\left(\frac{3}{\left(kR\right)^2}-1\right)-\frac{3\cos{\left(kR\right)}}{\left(kR\right)^2}\right]\mathbf{e_\rho}\right\};	
	\end{aligned}
	\label{EqEB}
\end{equation}
where $P_{\mathrm{PW}}=P/(1~\mathrm{PW})$, is the dimensionless wave power, $P$ is the wave power $F_0=\frac{2e}{mc^2}\sqrt{\frac{3}{c}10^{22}\mathrm{erg\cdot~s^{-1}}}\approx1174$, $e=4.8\times10^{-10}\mathrm{~statC}$ is the elementary charge, $m$ is the positron mass, $c$ is the light velocity, $t$ is time, $\omega=2.1\times10^{15}\mathrm{~s^{-1}}$ is the wave frequency corresponding to the wave period $T=3~\mathrm{fs}$ and the wavelength $\lambda=0.9~\mathrm{\mu m}$, $k=\omega/c$ is the wave number, $\rho$ and $z$ are radial and axial coordinates in the cylindrical coordinate system, $R=\sqrt{\rho^2+z^2}$, $\mathbf{e_\mathrm{\rho,\varphi,z}}$ are unit vectors. Fields are normalized to the relativistic value $mc\omega/e$. Distributions of electric and magnetic fields are shown in Fig.~\ref{FigFields}.

The characteristic field extrema are shown in Fig.~\ref{FigFields}(b). The maximum amplitude of the magnetic field $a_B=a$ is achieved at the central point $\rho=0,~z=0$ and the maximum amplitude of the electric field $a_E$ is achieved on a circle lying in the central plane $z=0$ with radius $\rho_m=0.33\lambda$. The dependencies of field amplitudes on the dimensionless power $P_\mathrm{PW}$ are as follows:
\begin{equation}
	\begin{aligned}
		a_B=&~a=2F_0\sqrt{P_\mathrm{PW}}/3\approx780\sqrt{P_\mathrm{PW}};\\
		a_E=&~0.65a\approx510\sqrt{P_\mathrm{PW}}.
	\end{aligned}
	\label{EqEBamp}
\end{equation}

The first node of the electric field coincides with the $z$ axis and the second electric field node lies approximately on the sphere with the radius of $0.72\lambda$. Near the central point electric and magnetic fields can be approximated as

\begin{equation}
	\begin{aligned}
		\mathbf{E}\approx&~ ak\rho\left(\frac{1}{2}-\frac{\left(kz\right)^2}{20}\right)\mathbf{e_\varphi};\\
		\mathbf{B}\approx&~ a\left\{\left[\left(1-\frac{\left(kz\right)^2}{10}\right)+\left(k\rho\right)^2\left(\frac{\left(kz\right)^2}{70}-\frac{1}{5}\right)\right]\mathbf{e_\mathit{z}}+\frac{k^2z\rho}{10}\mathbf{e_\rho}\right\}.
	\end{aligned}
	\label{EqEBapp}
\end{equation}

\section{Modes of particle motion\label{Sec3}}

\subsection{General view\label{Sec3_1}}

One of the main objectives of this work is to classify the possible regimes of particle motion depending on the power of the incident waves, and select the most characteristic ones which either can determine the further dynamics of vacuum breakdown or possess new fundamental properties, including those useful for possible applications. In the fields of tightly focused waves, the motion of particles, which may turn out to be essentially three-dimensional, strongly depends on the initial conditions: particle position, its momentum, initial phase of fields and so on (within this work the term particle refers only to electrons and positrons). Such a multidimensional problem is challenging not only for analytical consideration but even for numerical simulations.

To get started, we look at the general properties which can be set analytically. First, the motion is actually three-dimensional as opposed to the case of an e-dipole wave, where the particle trajectories mainly lie in the plane ($z$, $\rho$) since the magnetic field only has an azimuthal component and there is no force acting on the particle in the azimuthal direction \cite{Gonoskov_PRL2014,Bashinov_QE2013}. In the case of interest, the deflection from planar motion occurs because the electric field drives the particle in the azimuthal direction, while the magnetic part of the Lorentz force excites radial and axial motion.

Second, particles oscillating along the azimuth experience centrifugal force. This force is represented by the second term in the radial projection of the motion equation in the cylindrical coordinate system, while the first term $F_{L\rho}$ is the radial projection of the Lorentz force:
\begin{equation}
	\frac{dp_\rho}{dt}=F_{L\rho}+\frac{p_\varphi^2}{m\gamma\rho},
	\label{EqMotionr}
\end{equation}
where $p_\rho$ and $p_\varphi$ are the radial and azimuthal components of particle momentum, $m$ and $\gamma$ are the particle mass and the Lorentz factor. Note, that the centrifugal force is a feature of the field configuration with an azimuthal electric field. It is absent in fields of the e-dipole wave where motion is radial and axial.

Third, in the ultrarelativistic case $\gamma\gg 1$ particle motion without photon recoil is almost independent of the wave amplitude. Indeed, equations of motion in variables $\mathbf{p}/a$ and $\mathbf{r}$ depend only on field structure, where $\mathbf{p}$ and $\mathbf{r}$ are the particle momentum and the position vector. So, in the case of photon recoil taken into account any difference in the character of motion is an evidence of radiative effects.

For modeling of particle motion, we use the QED-PIC code PICADOR \cite{Surmin_CPC2016} simulating photon emission by particles as random acts in frame of the quasiclassical approach \cite{BayerKatkov} with help of the Monte-Carlo method \cite{Gonoskov_PRE2015}. It also has an option which allows excluding radiative recoil in order to simulate motion of particles which can emit photons but experience only the Lorentz force. This option is very useful to compare results of simulations with and without radiation losses. In order to identify particle motion and retrieve different characteristics of dynamics of the particle ensemble, in the next sections we consider the following numerical setup. Test electrons and positrons are initially distributed uniformly within a sphere with the radius of $0.6\lambda$ and the center coinciding with the central point. The initial number of particles of each type is $3\times10^6$, fields are set analytically according to Eq.~\ref{EqEB}. The size of the simulation box is $6\lambda\times6\lambda\times6\lambda$ along $x$, $y$ and $z$ axes and the number of cells along them is $600\times600\times600$. The time step is $T/260$. The wave power varies from 0.01~PW to 300~PW.

\subsection{Particle escape\label{Sec3_2}}

\begin{figure}
	\includegraphics[width = 1\columnwidth]{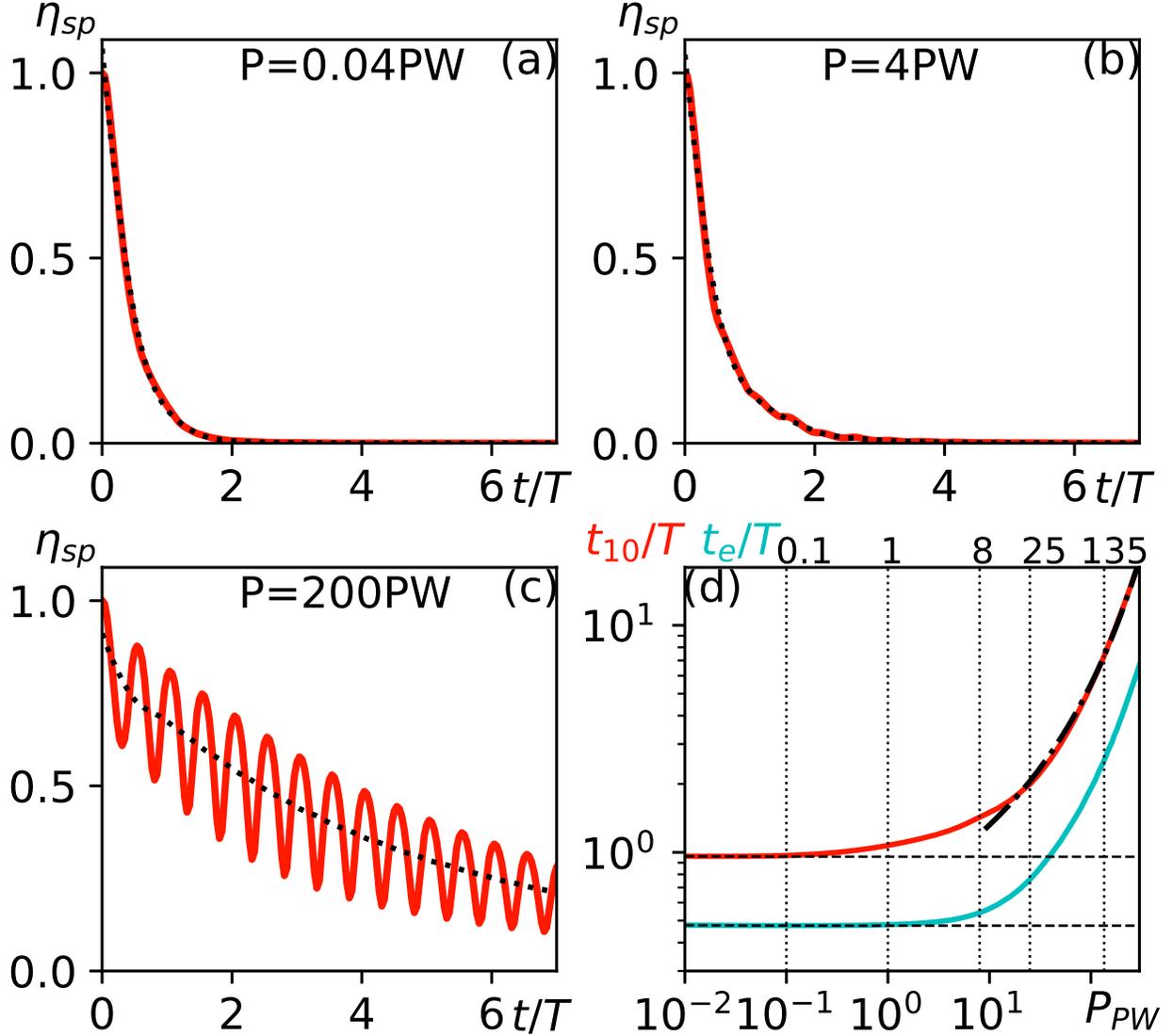}
	\caption{(Color online) Temporal evolution of particle number in a centered sphere with the radius of $0.6\lambda$ normalized to its initial value $\eta_{\mathrm{sp}}(t)=N_{\mathrm{sp}}(t)/N_{\mathrm{sp}}(0)$ for different wave powers: (a) $P=0.04~\mathrm{PW}$; (b) $P=4~\mathrm{PW}$; (c) $P=200~\mathrm{PW}$. Red (dark gray) solid lines correspond to primary data of numerical simulations, black dotted lines are results of their smoothing by Savitzky-Golay filter.(d) Dependencies of time intervals $t_{10}$ (red (dark gray) solid line) and $t_e$ (pale blue (gray) solid line) on the wave power. For comparison the corresponding time intervals obtained without photon recoil are shown by black dashed lines. Dash-dotted line shows an approximation of $t_{10}$ at great powers. Vertical dashed lines correspond to threshold powers of different regimes of motion.\label{FigEscape}}
\end{figure}

First of all, we would like to point out that the difficulties of analysis of particle motion is particle escape from focus due to strong field inhomogeneity and stochasticity of photon emission. The stationary particle distributions giving the most simple and rigorous criteria of different modes of motion are not formed as in the case of a plane wave \cite{Gonoskov_PRL2014,Kirk_PPCF2016,Bashinov_PRA2017,Bulanov_JPL2017,Bashinov_QE2018}. However, temporal evolution of the particle ensemble in the focus, its spatial distribution at particular moments of time as well as energy and angular distribution of particles and generated photons have imprints of specific features of motion and provide an opportunity to distinguish different regimes of interaction \cite{Bashinov_QE2019}.

After the start of movement particles inevitably leave the focal region, so the particle number $N_{\mathrm{sp}}$ within the initial sphere decreases. At powers $P\lesssim1~\mathrm{PW}$ $N_{\mathrm{sp}}(t)$ decreases monotonically (see Fig.~\ref{FigEscape}(a)), while at greater powers $N_{\mathrm{sp}}(t)$ exhibits modulations at the doubled wave frequency as shown in Fig.~\ref{FigEscape}(b),~(c). The modulations are results of radiation losses, because without photon recoil within the considered power range $N_{\mathrm{sp}}(t)$ is the same as one at $P\lesssim1~\mathrm{PW}$ as simulations confirm.

To retrieve the characteristic time of change of $N_{\mathrm{sp}}(t)$ the Savitzky-Golay filter \cite{SavGol_AC1964} is applied. This filter smoothes modulations and allows obtaining the average particle number $\overline{N}_{\mathrm{sp}}(t)$ within the initial sphere. A common characteristic time of particle escape $t_e$ defined as $\overline{N}_{\mathrm{sp}}(t_e)=N_{\mathrm{sp}}(0)/e$ is shown in Fig.~\ref{FigEscape}(d). On one hand it characterizes radiative effects since it also indicates $P\approx1~\mathrm{PW}$ as a threshold power, starting from which $t_e$ increases. On the other hand particles may need a longer time than $t_e$ to accumulate radiative effects and begin to oscillate in a quasi steady-state radiative regime. For this reason we introduce another characteristic $t_q$ as $\overline{N}_{\mathrm{sp}}(t_q)=N_{\mathrm{sp}}(0)/q$, where $q$ is a certain number. The larger $q$, the larger $t_q$, but we found that the form of $t_q(P)$ almost stops changing at $q>7$. So, instead of $t_e$ the time interval $t_{10}$ is more sensitive to radiative effects which emerge even at $P\gtrsim0.1~\mathrm{PW}$ according to $t_{10}$ (compare red solid and dashed black lines in Fig.~\ref{FigEscape}(d)).

Thus, analysis of character of particle escape reveals two threshold powers $P_{th1}\approx0.1~\mathrm{PW}$ and $P_{th2}\approx1~\mathrm{PW}$ which indicate changes in the evolution of particle distribution. These thresholds are marked by vertical dashed lines in Fig.~\ref{FigEscape}(d). However, at greater powers, when radiation losses should become more and more prominent, dependencies $t_{10}(P)$ and $t_{e}(P)$ do not allow distinguishing any evident thresholds. These thresholds (vertical dashed lines at $P>P_{th2}$ in Fig.~\ref{FigEscape}(d)) will be obtained in the next section.

\subsection{Spatial distributions, energy, and angular spectra of particles\label{Sec3_3}}

Since without photon recoil in the relativistic case particle motion does not depend on wave power, the spatial distribution of particles and their distribution in momentum space with respect to $\mathbf{p}/a$ should demonstrate clearly radiative effects. In this section based on proposed distributions we determine characteristic power thresholds of different regimes of motion.

\begin{figure}
	\includegraphics[width = 1\columnwidth]{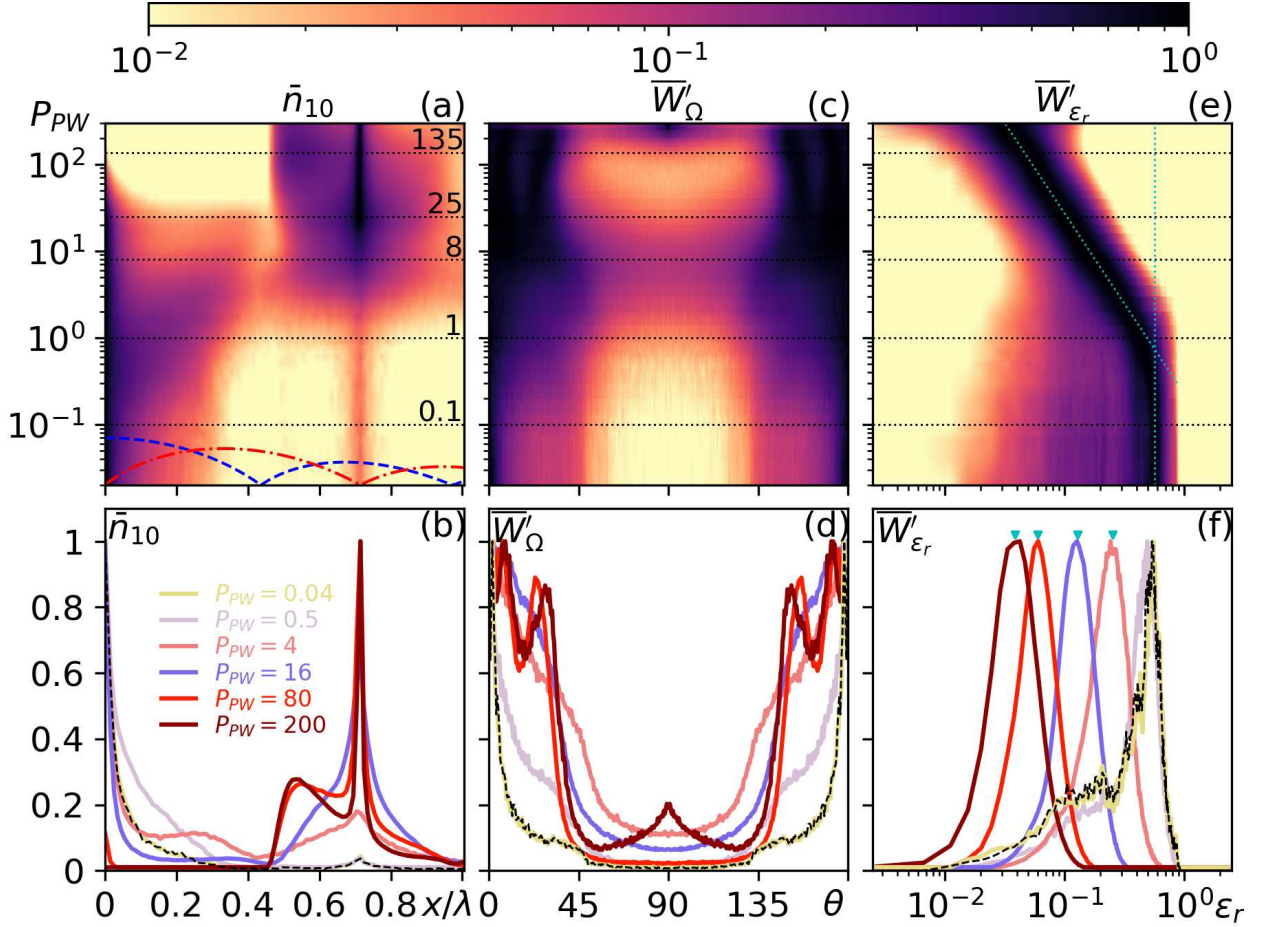}
	\caption{(Color online) (a) Map of particle distribution in radial direction $\overline{n}_{10}(\rho,\varphi_0=0)\approx\overline{n}_{10}(x)$ as function of wave power. Blue (black) dashed and red (gray) dash-dotted lines correspond to distributions of magnetic and electric field amplitudes along the radial direction in the central plane. Maps of angular $\overline{W}_{\Omega}^{'}$ and energy $\overline{W}_{\varepsilon_r}^{'}$ distributions of particles passing the observation sphere as functions of wave power are presented in figures (c) and (e), respectively. Horizontal black dotted lines show power threshold of different dynamics of the particle ensemble. Vertical and sloped cyan (gray) lines correspond to approximations of maxima locations $\varepsilon_{r,m}$ of $\overline{W}_{\varepsilon_r}^{'}$: $\varepsilon_{r,m}\approx0.6$ and $\varepsilon_{r,m}\approx0.5P_{\mathrm{PW}}^{-0.48}$, respectively. Locations of maxima of $\overline{W}_{\varepsilon_r}^{'}$ according to the last approximation are depicted by cyan (gray) arrows in the figure (f). Examples of averaged distributions are depicted in the figures (b), (d) and (f) by colored lines (in shades of gray) corresponding to different regimes of motion. For comparison, distributions obtained without photon recoil are shown by black dashed lines.\label{FigMaps}}
\end{figure}

The spatial particle distribution is one of the base characteristics of the particle ensemble. As was stated above, particle motion in the fields of the m-dipole wave is three-dimensional and a scan over the wave power adds a fourth parameter to analysis. However, the number of dimensions can be reduced. First, the axial field symmetry leads to a uniform particle distribution along the azimuth. Second, it is reasonable to consider particles in the vicinity of the central plane, because they need more time to escape the focal region and, as a consequence, the influence of radiative losses on these particles is more prominent. So the scan of the radial particle distribution over the wave power has to be quite sensitive to the radiative effects.

Since simulations are performed in Cartesian coordinates, due to axial symmetry the particle distribution is considered along the $x$ axis at $x>0$ without loss of generality. This distribution corresponds to the radial distribution at an azimuth $\varphi_0=0$. To retrieve the characteristic particle distribution from simulations the following procedure is carried out. First, the total number of particles in the vicinity of the $x$ axis in ranges $z\in[-\vartriangle_z,\vartriangle_z]$ and $y\in[-\vartriangle_y,\vartriangle_y]$ is calculated as function of time $N(t)=\int_{x=0}^\lambda\int_{z=-\vartriangle_z}^{\vartriangle_z}\int_{y=-\vartriangle_y}^{\vartriangle_y} n(x,y,z,t)d^3r$, where $n(x,y,z,t)$ is the particle density. Second, the averaged particle density along the $x$ axis is retrieved: $n_{av,q}(x)=\frac{1}{sT}\int_{t=t_q}^{t_q+sT}\int_{z=-\vartriangle_z}^{\vartriangle_z}\int_{y=-\vartriangle_y}^{\vartriangle_y}\frac{n(x,y,z,t)}{N(t)}dtdzdy$, where $s$ is an integer number. The normalization is needed because the particle number in the focus decreases in time (Fig.~\ref{FigEscape}), and otherwise only initial moments of time may contribute to the spatial distribution significantly. Finally, we obtain the characteristic spatial particle distribution as $\overline{n}_q(x)=n_{av,q}(x)/n_{av,q}^{\mathrm{max}}(x)$ and $\overline{n}_q(x)$ is approximately $\overline{n}_q(\rho,\varphi_0)$. In order to decrease computational noises without introducing artificial effects the following parameters were chosen: $s=2;~\vartriangle_z=0.1\lambda; \vartriangle_y=0.04\lambda$. $\overline{n}_q(x,P)$ depends on $q$, however, as in case of $t_q(P)$, at $q>7$ the characteristic distribution is stabilized and the map of distributions $\overline{n}_{10}(x,P)$ is considered (Fig.~\ref{FigMaps}(a),(b)).

Apart from particle distribution in the focal region (which is quite difficult to obtain in experiments), from a practical point of view angular distributions and energy spectra of particles escaping the focal region are of significance. By analogy with the averaged spatial distribution, averaged angular and energy characteristics can be introduced. The difference is that these characteristics are measured at a distance from the focus: at the surface of an observation sphere with radius $r_\mathrm{obs}$ which is equal to half of the size of the simulation box. Thus, an event in focus can be observed via these characteristics approximately $t_d=r_\mathrm{obs}/c$ later, in our simulations $t_d\approx3T$. 

Let $I_\Omega^{'}\equiv dI/d\Omega$ denote the fraction of energy $dI$ of particles passing through the observation sphere per unit time with momentum directed into an element of solid angle $d\Omega$ in the momentum space. Note that the angular distribution in momentum space measured in the micron-sized region is relevant to the angular distribution in coordinate space at much larger distance, where the experimental detectors are usually placed. The normalizing factor is equal to $I_{\Omega,n}=\int I_\Omega^{'}d\Omega$. The normalized energy of particles per solid angle over time period $sT$ is $W_\Omega^{'}=1/\left(sT\right)\int_{t=t_q+t_d}^{t_q+t_d+sT}I_\Omega^{'}(t)/I_{\Omega,n}(t)dt$. Finally, the averaged angular characteristic of particles is $\overline{W}_\Omega^{'}=W_\Omega^{'}/W_\Omega^{'\mathrm{max}}$. Suitable parameters corresponding to stabilized angular distributions are the same as those for the spatial distribution: $s=2;~q=10$. The distribution of $\overline{W}_\Omega^{'}$ is uniform along azimuth owing to field structure and only its dependence on the polar angle $\theta$ (measured from the positive direction of the $z$ axis) is considered (Fig.~\ref{FigMaps}(c),~(d)).

Opposite to angular and spatial characteristics, an energy spectrum depends on power in the relativistic case without photon recoil. However, if $\varepsilon$ is the particle energy, the spectrum as a function of
\begin{equation}
	\varepsilon_r=\varepsilon/(mc^2a_E)
	\label{EqEnRel}
\end{equation}
(relative spectrum) maintains its form at different relativistic powers without photon recoil and can be considered as a sensitive characteristic of radiative effects. Following $\overline{W}_\Omega^{'}$ the averaged relative energy characteristic $\overline{W}_{\varepsilon_r}^{'}$ can be introduced by replacing $\Omega$ for $\varepsilon_r$ in the previous paragraph, and the suitable parameters are the same: $s=2;~q=10$ (Fig.~\ref{FigMaps}(e),~(f)).

Based on the introduced characteristics we reveal manifestations of radiative effects at different wave powers (Fig.~\ref{FigMaps}). In the region $P<P_{th1}$ these characteristics are almost independent of power and they are very close to those obtained without photon recoil, so radiative effects are negligible. In the vicinity of the central plane the major part of particles are within $\rho<\rho_m=0.33\lambda$ and form a peak of density at the first electric field node (the central point) (Fig.~\ref{FigMaps}(a),~(b)). Also there is a much weaker peak of particle density at the second electric field node ($\rho\approx0.72\lambda$) and there is almost no particles between the first electric field antinode ($\rho\approx\rho_m$) and the second electric field node. So, particles oscillate around the electric field nodes and we call this regime {\it ponderomotive trapping}. Trapped particles escape the focal region close to the $z$ axis ($\theta\approx0^{\circ},~180^{\circ}$) with an angular spread around $5^{\circ}$ and there is almost no transverse particle escape (Fig.~\ref{FigMaps}(c),~(d)). Since radiation losses are negligible, the majority of particles can gain high energy around $\varepsilon_{r,m}\approx0.6$ and escape focus without losing a large part of their energy (Fig.~\ref{FigMaps}(e),~(f)). The relative energy spread is determined mainly by initial conditions and is relatively small. Note that in the considered power range most particles initially located outside a sphere with the radius of $\rho_m$ quickly escape the focal region within $t<t_{10}$ and they are not taken into account in the averaged characteristics presented in Fig.~\ref{FigMaps}. Their motion can be called {\it ponderomotive escape} because the direction of their escape is mainly determined by field gradients.

At greater powers $P_{th1}<P<P_{th2}$ the relative number of particles between the first node and antinode of the electric field increases due to radiative effects (Fig.~\ref{FigMaps}(a),~(b)). This leads to an increase of the angle of particle escape with respect to the $z$ axis (Fig.~\ref{FigMaps}(c),~(d)) and few particles can escape the focal region at large angles $60^\circ<\theta<120^\circ$. This also leads to an increase of the relative number of particles which can gain large energy in the region of strong electric field that is confirmed by shift of the left edge of $\overline{W}_{\varepsilon_r}^{'}$ to larger $\varepsilon_r$ in Fig.~\ref{FigMaps}(e). At the same time most of the particles start to experience noticeable radiation losses and the location of maximum of averaged relative energy distribution begins to shift to lower $\varepsilon_r$ from $\varepsilon_{r,m}\approx0.6$ (Fig.~\ref{FigMaps}(e),~(f)). This regime can be named as {\it anomalous radiative trapping} (ART) \cite{Gonoskov_PRL2014}. The threshold of this regime corresponds to relatively low field amplitudes in relativistic units: $a_B=250$ and $a_E=160$. Such low threshold values are due to extreme focusing and \textit{ponderomotive trapping} near the central point. Particles are confined for some time in the vicinity of the central point by the “well-like” ponderomotive potential and they can accumulate influence of radiative effects. In the e-dipole wave with a “hill-like” ponderomotive potential near the center particles quickly leave the focal region at such powers and ART is observed at much greater powers \cite{Bashinov_QE2019}.

At powers above $P_{th2}$ ART becomes more prominent. Particles are drawn into the region of the electric field antinode $\rho_m<\rho<0.45\lambda$ (where without photon recoil there are no particles), and the additional maximum of $\overline{n}_{10}$ appears within $0.05\lambda<\rho<0.45\lambda$ (Fig.~\ref{FigMaps}(a),~(b)). As a result, the portion of particles leaving focal region at large $\theta$ significantly increases (Fig.~\ref{FigMaps}(c),~(d)). At the same time at these powers the relative number of particles in the vicinity of the second electric field node sharply increases and the so-called regime of {\it normal radiative trapping} (NRT) \cite{Gonoskov_PRL2014,Lehmann_PRE2012,Kirk_PPCF2009} appears. A particle escaping from focus can emit a hard photon, significantly increasing curvature of motion, and then start to oscillate around the electric field node for some time. Due to hard photon emission the maximum of $\overline{W}_{\varepsilon_r}^{'}$ is shifted to lower relative energies $\varepsilon_r$ and quite a good approximation for the location of the maximum in a wide range of powers $P_{th2}<P<300~\mathrm{PW}$ can be found:
\begin{equation}
	\varepsilon_{r,m}\approx0.5P_{\mathrm{PW}}^{-0.48}
	\label{EqEnRelMax}
\end{equation}
according to Fig.~\ref{FigMaps}(e),~(f). This approximation is evidence that the energy of the majority of escaping particles $\varepsilon_m$ becomes almost independent of wave power allowing for Eqs. \ref{EqEBamp} and \ref{EqEnRel}. In the range $P_{th2}<P<300~\mathrm{PW}$ $\varepsilon_m$ is from $250mc^2$ to $350mc^2$ or from $130~\mathrm{MeV}$ to $180~\mathrm{MeV}$.

Qualitatively the set of regimes of motion do not change if power exceeds $P_{th3}\approx8~\mathrm{PW}$. However, the relative number of particles moving in the regime of ART noticeably decreases and NRT regime becomes the most prominent, this is shown by the reduction of $\overline{n}_{10}$ in the range $0.05\lambda<\rho<0.45\lambda$ and by its rise in the range $0.5\lambda<\rho<0.9\lambda$ (Fig.~\ref{FigMaps}(a),~(b)). Due to particle redistribution between the regimes of motion regions of polar angle $0^\circ<\theta<25^\circ$ and $155^\circ<\theta<180^\circ$ appear to be more populated (Fig.~\ref{FigMaps}(c),~(d)). Averaged energy distribution approximately keeps the relative width and $\varepsilon_{r,m}$ decreases according to Eq.~\ref{EqEnRelMax} (Fig.~\ref{FigMaps}(e),~(f)).

If the wave power exceeds $P_{th4}\approx25~\mathrm{PW}$, then the impact of ART regime on averaged spatial distribution in the region $0.05\lambda<\rho<0.45\lambda$ almost disappears, but in addition to {\it ponderomotive trapping}, ART and NRT the ART regime emerges in another region $0.45\lambda<\rho<0.6\lambda$. There are two distinct maxima of $\overline{n}_{10}$ at $\rho=0$ and $\rho\approx0.72\lambda$ and the region $0.45\lambda<\rho<0.6\lambda$ corresponding to the emerging ART regime becomes more populated by particles and at $P\approx70~\mathrm{PW}$ the third maximum at $\rho\approx0.55\lambda$ appears (Fig.~\ref{FigMaps}(a),~(b)). Particles in the second ART regime oscillate closer to the second electric field node than to the electric field antinode as it is in the case of the e-dipole wave \cite{Gonoskov_PRL2014}. As a result, the relative averaged spectrum does not change qualitatively and maintains the same behavior as for lower powers: the relative energy width is approximately constant and the location of maximum is described by Eq.~\ref{EqEnRelMax} (Fig.~\ref{FigMaps}(e),~(f)). However, there are changes in the averaged angular characteristic. First, $\overline{W}_\Omega^{'}$ becomes non-monotonous in the ranges $0^\circ<\theta<45^\circ$ and $135^\circ<\theta<180^\circ$, and in each region two maxima appear. Second, $\overline{W}_\Omega^{'}$ in the region $45^\circ<\theta<135^\circ$ significantly decreases (Fig.~\ref{FigMaps}(c),~(d)).

At powers greater than $P_{th5}\approx135~\mathrm{PW}$ the averaged spatial distribution does not demonstrate any changes except the near disappearance of the maximum in the center ($\rho=0$) corresponding to the {\it ponderomotive trapping regime} (Fig.~\ref{FigMaps}(a),~(b)), however angular and energy distributions differ qualitatively. A portion of the particles escape the focal region preferably transversely to the $z$ axis and one of the maxima of $\overline{W}_\Omega^{'}$ lies at $\theta=90^\circ$. At $P>250~\mathrm{PW}$ the transverse escape becomes primary. The averaged relative energy distribution becomes wider, which is especially clear at the tails of the distribution, but the location of the maximum is in accordance with the approximation~\ref{EqEnRelMax}. Reasons of such change will be analyzed in the next section.

Thus, introduced averaged characteristics can be considered as indicators of radiative effects. Joint analysis of these characteristics allows distinguishing five threshold powers which separate different dynamics of the particle ensemble. These changes can be caused by the emergence of new regimes of motion and the redistribution of particles between these regimes. Proposed characteristics provide a general view on particle motion at the “macrolevel” and the necessity of comprehension of particle trajectories remains.

\subsection{Analysis of particle trajectories\label{Sec3_4}}

Analysis of individual particle motion is necessary in order to understand observed macrocharacteristics and can help explain properties of different regimes hidden at the “macrolevel”. In this section we consider particle trajectories observed during evolution of ensembles of electrons and positrons described in Sec.~\ref{Sec3_3} in the distinguished ranges of the wave power. Initially particles at rest are distributed uniformly within a sphere with radius $0.6\lambda$ with the center in the central point. Fields are set analytically according to Eq.~\ref{EqEB}. Parameters of simulations are described in Sec.~\ref{Sec3_1}. Since in the previous section the spatial regions on both sides of the first electric field antinode have been distinguished, we pay special attention to particles initiated in these regions.

\begin{figure}
	\includegraphics[width = 1\columnwidth]{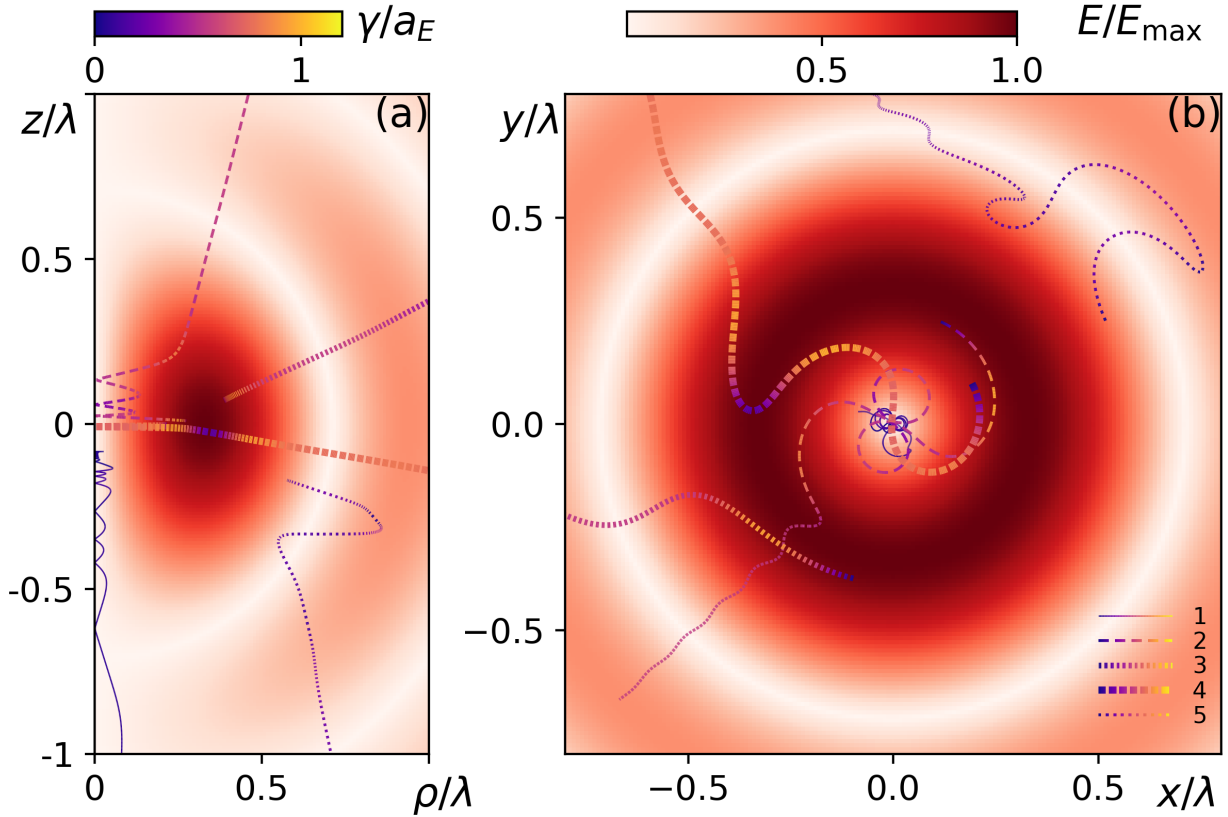}
	\caption{(Color online) Examples of trajectories in fields of the m-dipole wave with power of $0.04~\mathrm{PW}$ (a) in plane $z\rho$ and (b) projections of the trajectories onto the central $xy$ plane. Colors along trajectories denote the Lorentz factor values. The electric field distribution is shown  in shades of red (gray).\label{FigTraj004}}
\end{figure}

For the sake of clarity first we consider the case of relatively low wave power ($P=0.04~\mathrm{PW}$, Fig.~\ref{FigTraj004}) when radiative effects are negligible. In this case particle motion is mainly determined by field gradients. Particles within the region confined by the first electric field antinode (central region) oscillate in azimuthal and radial directions like in a ponderomotive potential and drift mainly along the z axis due to field inhomogeneity (trajectories 1 and 2 in Fig.~\ref{FigTraj004}). The drift can be changed by sudden escape if a particle is shifted from the central plane to the region of weaker fields and has enough energy to break out of {\it ponderomotive trapping} (trajectory 2 in Fig.~\ref{FigTraj004}).

Particles outside the central region quickly (within half a wave period) escape the focal region (trajectory 3 in Fig.~\ref{FigTraj004}). If a particle is located close to the second electric field node, it can be {\it ponderomotively trapped} in this region since it cannot gain sufficient energy in order to overcome the region of the next electric field antinode (trajectory 5 in Fig.~\ref{FigTraj004}). When trapped, a particle drifts along the electric field node in the direction of weaker fields where it can be released.

Thus presented trajectories completely explain the obtained average spatial distribution in the power range $P<P_{th1}$ (Fig.~\ref{FigMaps}(a),~(b)). A major portion of the particles oscillate within the central region where particles are {\it ponderomotively trapped}, {\it ponderomotive trapping} at the second electric field node is also observed, and the other regions are almost unpopulated by particles.

However it is necessary to mention one more type of trajectories: numerical results show that some particles initiated in the central region can leave it. The centrifugal force facilitates this escape. Let us suppose that for a particle at a certain distance from the central point the centrifugal force cannot be balanced by the Lorentz force, so this particle is pushed in the radial direction outwards from the central region (trajectory 4 in Fig.~\ref{FigTraj004}).

\begin{figure}
	\includegraphics[width = 1\columnwidth]{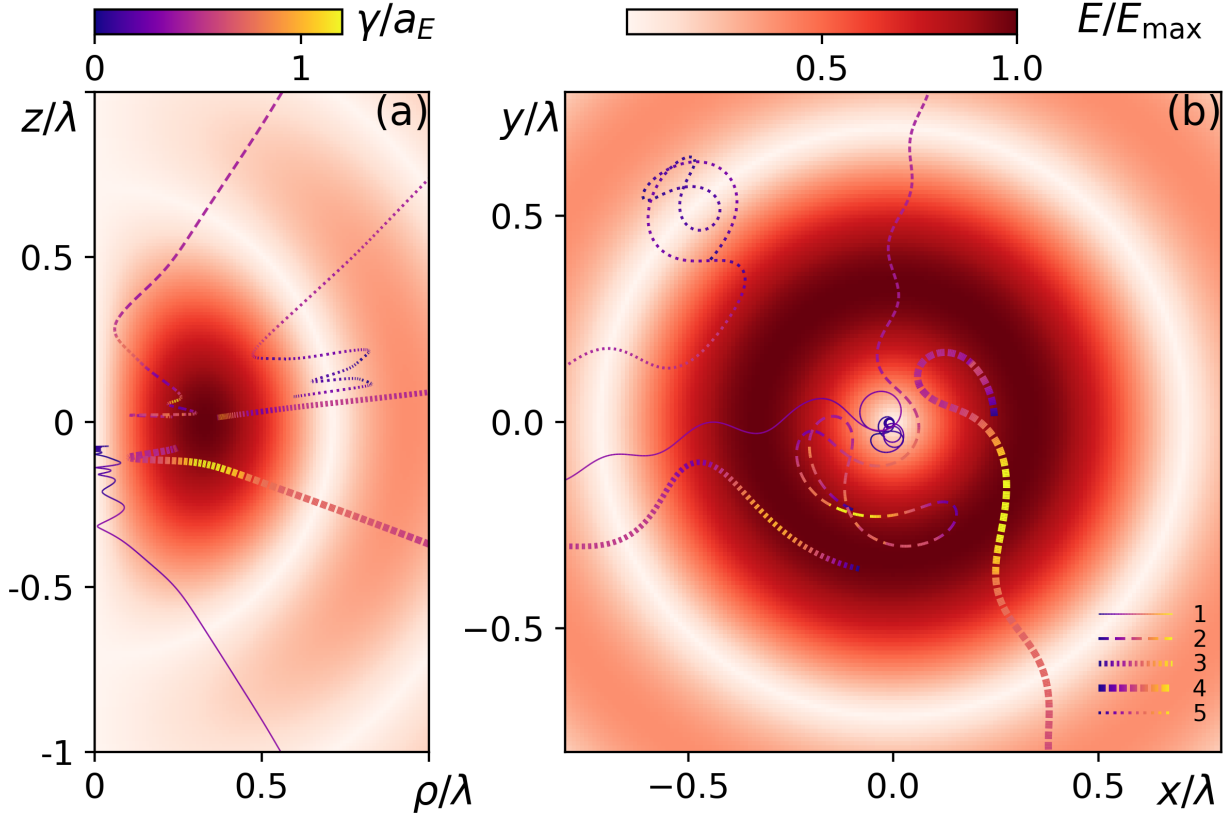}
	\caption{(Color online) Examples of trajectories in fields of the m-dipole wave with power of $0.5~\mathrm{PW}$ (a) in plane $z\rho$ and (b) projections of the trajectories onto the central $xy$ plane. Colors along trajectories denote the Lorentz factor values. The electric field distribution is shown  in shades of red (gray).\label{FigTraj05}}
\end{figure}

\begin{figure}
	\includegraphics[width = 1\columnwidth]{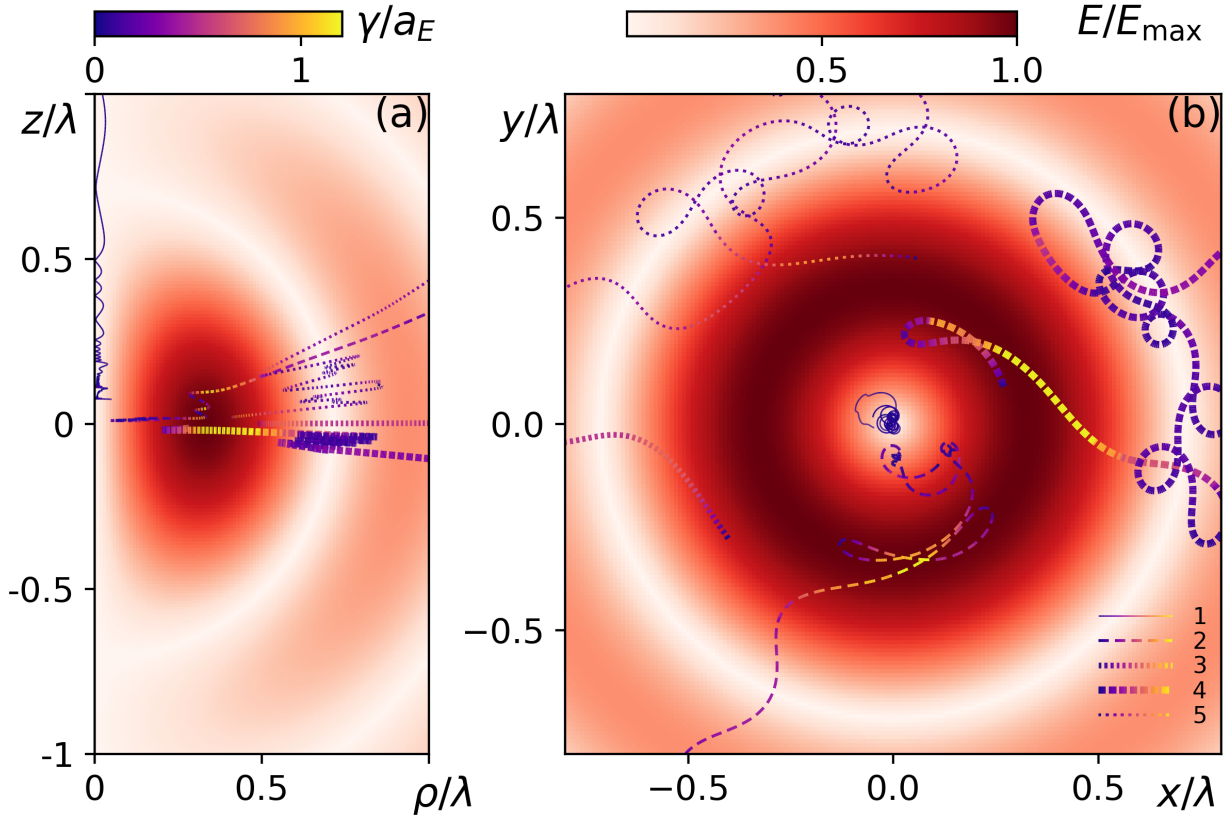}
	\caption{(Color online) Examples of trajectories in fields of the m-dipole wave with power of $4~\mathrm{PW}$ (a) in plane $z\rho$ and (b) projections of the trajectories onto the central $xy$ plane. Colors along trajectories denote the Lorentz factor values. The electric field distribution is shown  in shades of red (gray).\label{FigTraj4}}
\end{figure}

From Eq.~\ref{EqMotionr} one may obtain that the boundary of the balance is determined by
\begin{equation}
	m\omega B_z=p_\varphi/\rho
	\label{EqCentrifugBal}
\end{equation}
where the field is dimensionless. Since the maximal displacements of a particle from the $z$ axis during the wave period mainly occur when the magnetic field is close to its maximal value, the instantaneous field can be replaced by its local amplitude. At this moment the azimuthal electric field is changing its direction and the azimuthal momentum is close to its maximum value. Assuming that the maximum azimuthal momentum is a fraction $\alpha\in[0;1]$ of the maximal possible value $2mcE_m$ in a harmonic electric field, where $E_m$ is the local dimesionless field amplitude, using Eq.~\ref{EqEBapp} we arrive at the expression for the boundary radius in the vicinity of the central plane ($z\approx0$):
\begin{equation}
	\rho_b=\frac{\sqrt{5(1-\alpha)}}{2\pi}\lambda\approx0.36\sqrt{1-\alpha}\lambda.
	\label{EqBoundRho}
\end{equation}
In accordance with Fig.~\ref{FigMaps}(e) it is reasonable to consider $\alpha\approx0.3$ on average so $\rho_b\approx0.3\lambda$. Though it is quite a rough estimate, this value is in accordance with the averaged particle distribution (Fig.~\ref{FigMaps}(a),~(b)) which shows that largest portion of particles are within $\rho<\rho_b$ and particles crossing this boundary quickly escape (see also trajectories 4 in Fig.~\ref{FigTraj004}-Fig.~\ref{FigTraj4}).

Also the presented trajectories at powers $P<P_{th1}$ explain the averaged angular and energy distributions obtained from numerical simulations. After $t_{10}\approx1T$ at considered powers (Fig.~\ref{FigEscape}(d)) particles remaining in the focal region are mainly {\it ponderomotively trapped} and they eventually escape this region close to the $z$ axis. This corresponds to Fig.~\ref{FigMaps}(c),~(d). Since trapped particles oscillate around the first node of the electric field, most of them cannot be accelerated by the strongest electric field, and the escaping particles’ energy is less than $a_E mc^2$. This result is confirmed by Fig.~\ref{FigMaps}(e),~(f).

At powers $P_{th1}<P<P_{th2}$ trajectories (Fig.~\ref{FigTraj05}) are similar to those in the previously considered power range, however, the influence of radiative effects on particle motion becomes noticeable. Some of the particles initiated close to the $z$ axis can be deflected to the region of strongest electric field (trajectories 1 in Fig.~\ref{FigTraj05}), while the particles initiated closer to the first electric field antinode can be retained there for several halves of wave periods before escaping (trajectories 2 and 4 in Fig.~\ref{FigTraj05}). On average in this power range the time of escape only slightly exceeds $1T$, because the drift along the $z$ direction is sufficiently fast. These particles prior to escape move in the ART regime, note that projections of their trajectories onto the $xy$ plane correspond to the eight-like motion as in the e-dipole wave at powers $P>10~\mathrm{PW}$ \cite{Bashinov_QE2013,Gonoskov_PRX2017}. In the ART regime particles are accelerated mainly along the electric field (along azimuth) and when the magnetic field exceeds the electric field, particles make a turn, losing a great part of the gained energy due to photon emission, and after that are accelerated again along the electric field but in the opposite azimuthal direction. Note that during the time of trapping radiative effects nearly compensate the impact of the centrifugal force. Emerging of the ART regime also explains the growth of the averaged spatial distribution in the region $0.05\lambda<\rho<0.33\lambda$ (Fig.~\ref{FigMaps}(a),~(b)) and the slight growth of the relative number of highly energetic particles (Fig.~\ref{FigMaps}(e),~(f)). Deflection of particles from the $z$ axis and their oscillations at a distance from it lead to accumulation of influence of field inhomogeneity and favor particle escape at larger angles with respect to the $z$ axis (trajectories 1 and 2 in Fig.~\ref{FigTraj05} and Fig.~\ref{FigMaps}(c),~(d)). On the contrary, in the case of the e-dipole wave ART changes transverse particle escape to axial particle escape due to the different field structure \cite{Bashinov_QE2019}.

Escape of particles outside the central region (trajectories 3 in Fig.~\ref{FigTraj05}) and {\it ponderomotive trapping} in the second electric field node region (trajectories 5 in Fig.~\ref{FigTraj05}) are almost the same as at lower powers.

If power becomes larger ($P_{th2}<P<P_{th3}$) radiative effects become more prominent. First, the region of {\it ponderomotive trapping} in the vicinity of the $z$ axis becomes narrower (trajectory 1 in Fig.~\ref{FigTraj4}) and the region of {\it anomalous trapping} grows. Out of particles initially located close to the axis, more drift in the radial direction and get attracted into the strongest electric field region, moving in the ART regime (trajectory 2 in Fig.~\ref{FigTraj4}). At the same time they gradually drift outwards in the axial direction. The radial drift causes an increase of the averaged density in the strong electric field region (Fig.~\ref{FigMaps}(a),~(b)). When particles are shifted due to axial drift by approximately the distance equal to the scale of electric field inhomogeneity along the $z$ axis, trapping is broken and they can escape at large angles to the $z$ axis. On average the escape occurs in $1-1.5T$. Due to the random nature of photon emission the centrifugal force may not be balanced by radiation losses and a particle can be pushed out of the central region (trajectories 2 and 4 in Fig.~\ref{FigTraj4}) within $0.5T$.

Second, particles can be trapped in the region of the second electric field node due to radiative effects. Opposite to {\it ponderomotive trapping} in this region when only closely located particles are trapped, {\it radiative trapping} means that even highly energetic particles, starting near the antinode, can become trapped after reaching the strong magnetic field region and emitting a significant part of their energy (trajectories 4 and 5 in Fig.~\ref{FigTraj4}). This regime is NRT, because particles are trapped in the same region as {\it ponderomotively trapped} particles, but as a result of intensified photon emission. When trapped, particles drift along the sphere corresponding to the second electric field node and gyrate in the strong magnetic field. The radius of rotations oscillates depending on the instantaneous magnetic field and the accumulated energy. However, radiation losses in the power range $P_{th2}<P<P_{th3}$ are not strong enough to keep the trapped particles in the electric field node for longer than $3T$. Although the relative number of particles in the NRT regime increases (Fig.~\ref{FigMaps}(a),~(b)), most particles initiated out of the central region escape within $0.5T$ without trapping (trajectory 3 in Fig.~\ref{FigTraj4}). Note, that escaping particles not trapped in the NRT regime, when passing the region of NRT regime, can emit photons, so the maximum of the averaged energy distribution is shifted to lower energies (Fig.~\ref{FigMaps}(e),~(f)).

Third, due to radiative effects the axial drift slows down significantly, which is clear from comparison of Figs.~\ref{FigTraj05} and ~\ref{FigTraj4}. Mainly this fact leads to an increase of time of escape from the trapped state (Fig.~\ref{FigEscape}) and favors transverse escape with respect to the $z$ axis (Fig.~\ref{FigMaps}(c),~(d)).

Fourth, in the considered power range the quantum nature of photon emission becomes crucial. This is determined by the dimensionless quantum parameter \cite{Nikishov_85,Ritus_85}
\begin{equation}
	\chi=\eta\sqrt{\left(\varepsilon\mathbf{E}/\left(mc^2\right)+\left[\mathbf{p}/(mc)\times\mathbf{B}\right]\right)^2-\left(\mathbf{p}/(mc)\cdot\mathbf{E}\right)^2},
	\label{EqChi}
\end{equation}
where $\eta=\hbar\omega/\left(mc^2\right)\approx2.7\times10^{-6}$ and fields are dimensionless. This parameter can be largest along the trajectories of particles which gain energy in the region of strong electric field and then turn in a strong magnetic field region or crossing this region (for example, trajectories 2-5 in Fig.~\ref{FigTraj4}). Indeed, assuming an accumulated energy of about $a_E mc^2$ and a strong magnetic field of about $0.34a$ (Fig.~\ref{FigFields}(b)), the maximal value of $\chi$ is about
\begin{equation}
	\chi_\mathrm{max}\approx0.34\eta a_E a\approx0.4P_\mathrm{PW}.
	\label{EqChiMax}
\end{equation}

Since in the considered power range the quantum parameter can be of the order of unity or even exceed it, not only the quantum correction of radiated power \cite{Berestetskii} but also stochasticity of photon emission becomes significant. In particular this results in randomness of trapping. One example of this is motion of particles initiated at similar initial positions and represented by trajectories 3 and 5 in Fig.~\ref{FigTraj4}.

\begin{figure}
	\includegraphics[width = 1\columnwidth]{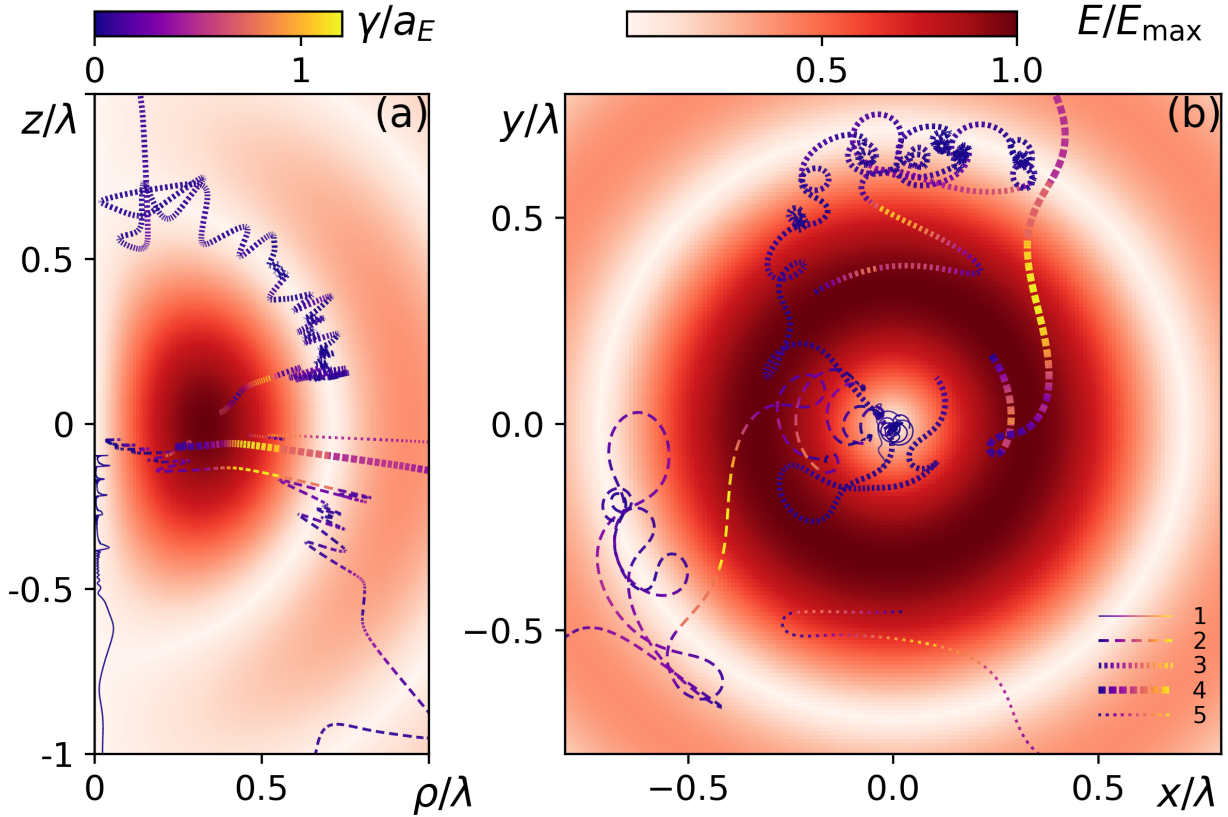}
	\caption{(Color online) Examples of trajectories in fields of the m-dipole wave with power of $16~\mathrm{PW}$ (a) in plane $z\rho$ and (b) projections of the trajectories onto the central $xy$ plane. Colors along trajectories denote the Lorentz factor values. The electric field distribution is shown  in shades of red (gray).\label{FigTraj16}}
\end{figure}

If the wave power is in the range $P_{th3}<P<P_{th4}$, possible regimes of motion remain the same. However, there are several important differences. The region of {\it ponderomotive trapping} near the $z$ axis begins to shift outwards from the central plane (trajectory 1 in Fig.~\ref{FigTraj16}), because ART becomes the main regime of motion in the vicinity of the central point (trajectory 2 in Fig.~\ref{FigTraj16}). If particles are initiated close to the central point, they move in the ART regime in the considered power range for up to $3T$ (trajectory 2 in Fig.~\ref{FigTraj16}), however in the case when the starting point is close to the electric field antinode particles on average need less than $1T$ to escape the central region (trajectory 4 in Fig.~\ref{FigTraj16}). In the antinode region particles can obtain a large azimuthal momentum and the centrifugal force increases the probability of their escape.

Comparison of trajectories 2 and 4 also reveals quantum stochasticity due to random acts of photon emission. Particles are accelerated up to similar energies but the first particle losing a larger part of its energy becomes trapped in the NRT regime while the second particle passes the region of the second electric field node.

Another difference is that NRT becomes much stronger and many particles escaping the central region become trapped. This is in accordance with the large maximum of the averaged spatial distribution in Fig.~\ref{FigMaps}(a),~(b). Due to the drift along the surface of the sphere corresponding to the second electric field node particles can reach the $z$ axis at a distance of around $0.7\lambda$ from the central plane (trajectory 3 in Fig.~\ref{FigTraj16}). In this region particle motion is chaotic and the particle can escape this region at random polar angles in ranges $0^\circ<\theta<25^\circ$ and $155^\circ<\theta<180^\circ$. Thus such strong trapping determines that these angular regions become more prominent (see Fig.~\ref{FigMaps}(c),~(d)).

\begin{figure}
	\includegraphics[width = 1\columnwidth]{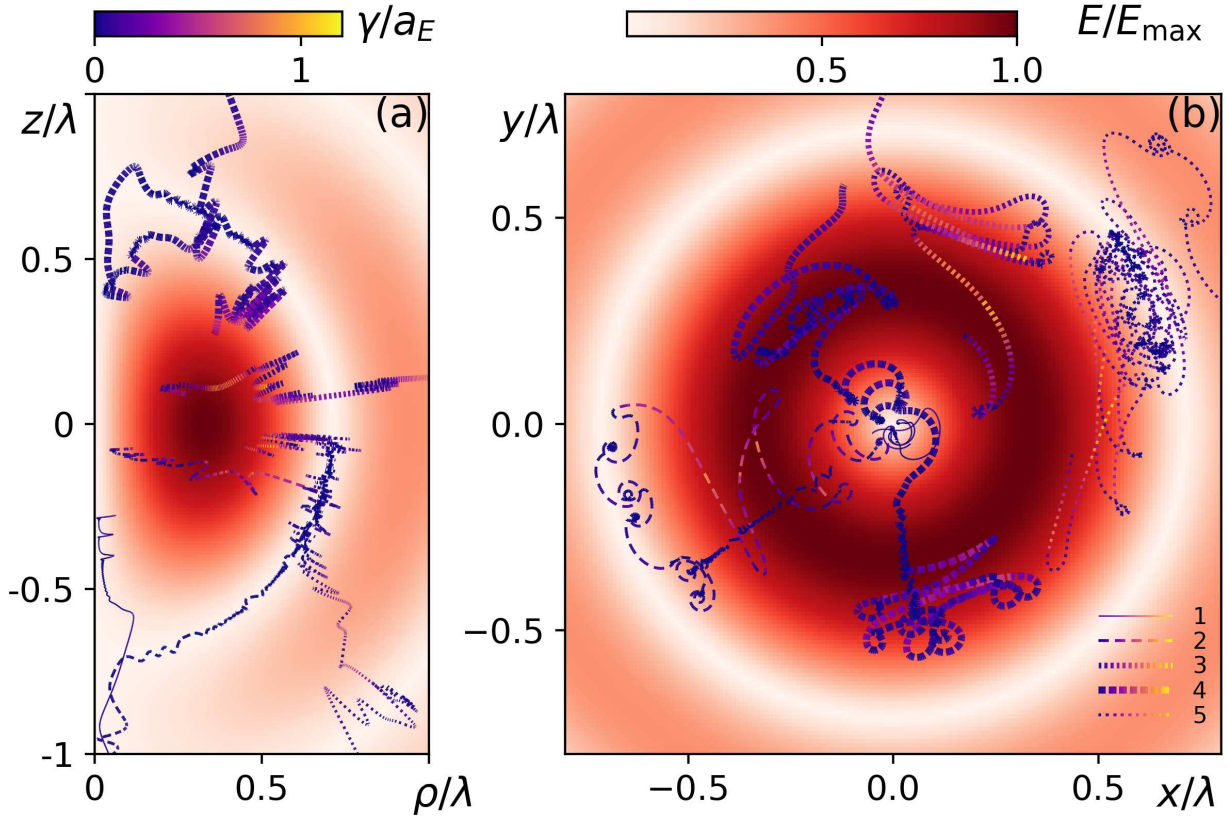}
	\caption{(Color online) Examples of trajectories in fields of the m-dipole wave with power of $80~\mathrm{PW}$ (a) in plane $z\rho$ and (b) projections of the trajectories onto the central $xy$ plane. Colors along trajectories denote the Lorentz factor values. The electric field distribution is shown  in shades of red (gray).\label{FigTraj80}}
\end{figure}

One more difference is related to the manifestation of trapping in farther regions. Particles can escape NRT in the region of the second electric field node due to attraction to the region of the second electric field antinode (in the ART regime) and after can be trapped in the region of the third node (in the NRT regime), see trajectory 2 in Fig.~\ref{FigTraj16}. However, these regimes are not as strong in more distant regions, so not many particles can be trapped there. Nevertheless, this contributes to the further decrease of relative energy of the escaping particles (Fig.~\ref{FigMaps}(e),~(f)).

Also a sign of a new regime of motion appears outside the central region between the first electric field antinode and the second electric field node (trajectory 5 in Fig.~\ref{FigTraj16}). A particle can spend about $1T$ in this region oscillating mainly along azimuthal direction instead of quick escape within $0.5T$ (compare trajectory 5 in Fig.~\ref{FigTraj16} with trajectories 3 in Figs.~\ref{FigTraj05} and \ref{FigTraj4}). In the power range $P_{th4}<P<P_{th5}$ such kind of motion transforms into the new regime. Particles escaping the central region can not only leave the focal region or become trapped in the vicinity of the second electric field node, but can also become trapped between the node and the first electric field antinode (trajectories 2-5 in Fig.~\ref{FigTraj80}). Although particles are trapped closer to the electric field node than to the antinode, this regime can also be called ART because such trapping contradicts the dynamics dictated by the ponderomotive potential.

The first ART regime emerging at $P>0.1~\mathrm{PW}$ is distinguished by radial attraction to the electric field antinode and axial drift outwards from the central plane, so we name it {\it radial} ART. In the second ART regime particles oscillate at distance around $0.55\lambda$ from the central point and drift towards the central plane, and we name this regime {\it axial} ART. This regime can follow {\it radial} ART, if a particle is pushed out of the central region by the centrifugal force (trajectory 2 and 3 in Fig.~\ref{FigTraj80}). Also particles can initially be in the basin of \textit{axial} ART (trajectory 4, 5 in Fig.~\ref{FigTraj80}). In the considered power range such trapping is no longer than $2-3T$ on average, which is sufficient to increase density and form a local maximum around $0.45\lambda<\rho<0.6\lambda$ of the averaged spatial distribution (Fig.~\ref{FigMaps}(a),~(b)). At the same time for many particles this trapping time is insufficient to closely approach the central plane (trajectory 3 and 4 in Fig.~\ref{FigTraj80}). It is more probable that a particle changes this regime to NRT in the vicinity of the second electric field node because NRT is very strong at the considered powers and can last up to $5T$. When moving in the NRT regime the particle drifts along the sphere, and can reach the $z$ axis with a high probability (trajectory 2 and 4 in Fig.~\ref{FigTraj80}) and escape the focal region at a polar angle $0^\circ<\theta<25^\circ$ or $155^\circ<\theta<180^\circ$. This explains significant particle escape at these angles and the reduction of transverse escape in Fig.~\ref{FigMaps}(c),~(d).

Note that {\it ponderomotive trapping} can occur mainly at large distance along the $z$ axis from the central point (trajectory 1 in Fig.~\ref{FigTraj80}), so the local maximum of the averaged spatial distribution at $\rho=0$ becomes small (Fig.~\ref{FigMaps}(a),~(b)). This regime almost completely gives way to {\it radial} ART in the vicinity of the central point. Although the imprint of {\it radial} ART in the averaged spatial distribution disappears at the considered powers (Fig.~\ref{FigMaps}(a),~(b)), particles can indeed move in this regime (trajectory 2 in Fig.~\ref{FigTraj80}), however, for most of them this regime is replaced by another within the time $t_{10}$ of particle escape from focus (Fig.~\ref{FigEscape}(d)).

{\it Radial} ART in the region of the second electric field antinode (trajectory 3, 5 in Fig.~\ref{FigTraj80}) and NRT in the region of the third node (trajectory 5 in Fig.~\ref{FigTraj80}) become prominent and the last one leads to further decrease of energy of escaping particles (Fig.~\ref{FigMaps}(e),~(f)). The emergence of NRT at different distances from the central point can explain the appearance of several local maxima in the averaged angular distribution of escaping particles within ranges of polar angle $0^\circ<\theta<25^\circ$ and $155^\circ<\theta<180^\circ$. However, this is beyond the scope of this paper.

\begin{figure}
	\includegraphics[width = 1\columnwidth]{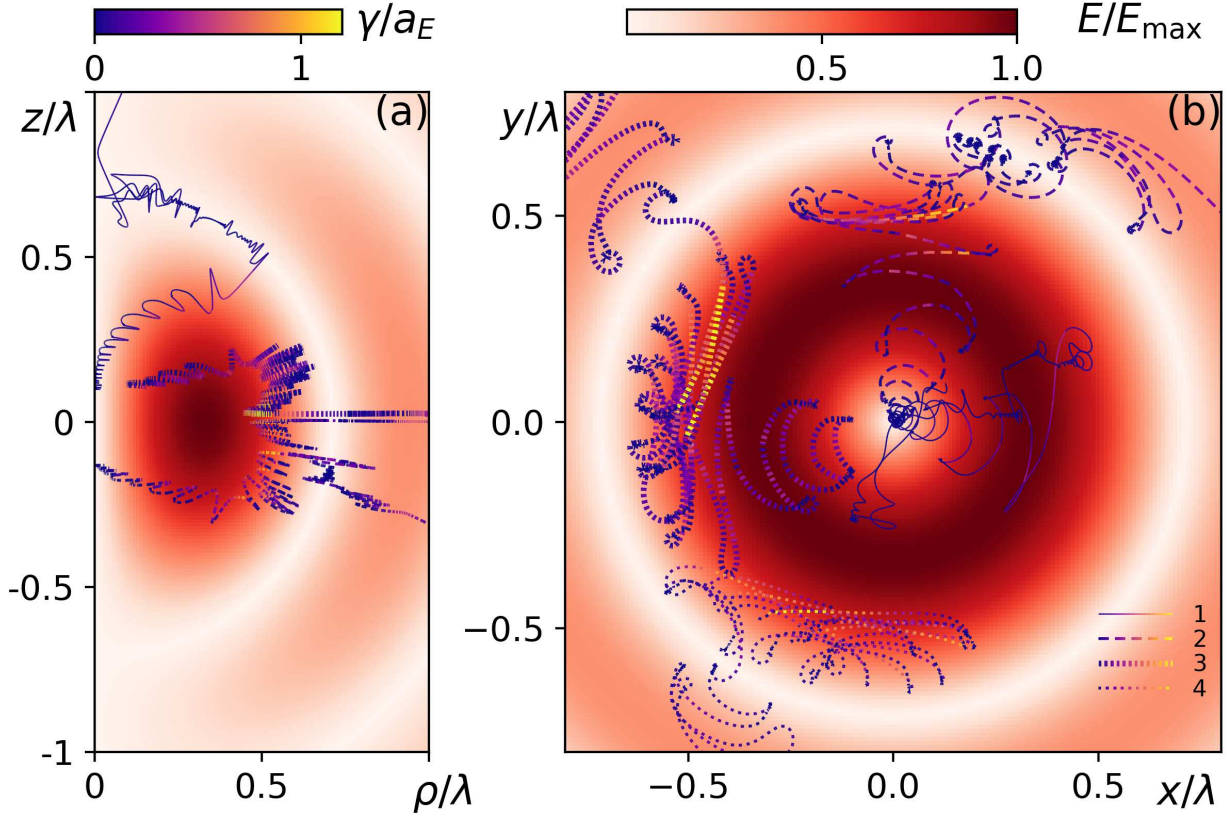}
	\caption{(Color online) Examples of trajectories in fields of the m-dipole wave with power of $200~\mathrm{PW}$ (a) in plane $z\rho$ and (b) projections of the trajectories onto the central $xy$ plane. Colors along trajectories denote the Lorentz factor values. The electric field distribution is shown  in shades of red (gray).\label{FigTraj200}}
\end{figure}

The key factor of particle motion in the power range $P_{th5}<P<300~\mathrm{PW}$ is the duration of motion in the {\it axial} ART regime. This duration becomes very long, reaching a few tens of wave periods, which allows most of the trapped particles to closely approach the central plane (trajectories 3 and 4 in Fig.~\ref{FigTraj200}). As a result the local maximum of the averaged angular distribution at $\theta=90^\circ$ appears (Fig.~\ref{FigMaps}(c),~(d)). The reasons are the following. First, the closer particles are to the central plane, the longer time is needed to move away from it after escaping the trapped region. Second, radiation losses significantly slow down the drift from this plane. Although after the {\it axial} ART regime particles can move in the NRT or the {\it radial} ART regimes, they usually do not obtain a large axial momentum and escape the focal region at large angles (trajectories 2-4 in Fig.~\ref{FigTraj200}). Transitions from {\it axial} ART to NRT do not allow reaching the absolute maximum of the averaged spatial distribution at $\rho\approx0.55\lambda$ and the absolute maximum remains at $\rho\approx0.72\lambda$ (Fig.~\ref{FigMaps}(a),~(b)).

Along with \textit{radial} ART in the region of the second electric field antinode and NRT in the region of the third electric field node, there is \textit{axial} ART in the region of the second electric field antinode, \textit{radial} ART in the region of the third electric field antinode (part of trajectory 3 in the left top corner of Fig.~\ref{FigTraj200}(b)) and NRT in the region of the fourth electric field node. Abundance of regions of NRT, together with stochasticity of photon emission, leads to a decrease of relative energy and quite a large energy spread of escaping particles (Fig.~\ref{FigMaps}(e),~(f)).

Escape of particles at small polar angles is ensured by escape from NRT like in the previous power range when particles drift in the region of the second electric field node to the axis (trajectory 1 in Fig.~\ref{FigTraj200}). However, at this power range the number of such particles decreases. Also \textit{ponderomotive trapping} was not observed at such powers in simulations. Particles initially located very close to the $z$ axis begin their motion in the \textit{radial} ART regime (trajectory 1 and 2 in Fig.~\ref{FigTraj200}).

Also we would like to emphasize the uniqueness of the \textit{axial} ART regime. If radiation losses are considered as a damping force, even with the quantum correction \cite{Kirk_PPCF2009} an attractor is formed in highly inhomogeneous fields of the m-dipole wave in the central plane at a distance of $\rho\approx0.55\lambda$. Thus only stochasticity of photon emission breaks the attractor in the \textit{axial} ART regime and limits trapping time. Note that in the case of the e-dipole wave although particles are attracted to the axis of symmetry in the ART regime, they eventually escape the focal region even if radiation losses are taken into account as a force.

Taking into account that the \textit{axial} ART regime essentially contributes to the escape time $t_{10}$ (Fig.~\ref{FigEscape}(d)) it is possible to fit the dependence $t_{10}(P)$. To remain trapped after acceleration by the electric field a particle needs to lose significant part of its energy each half of the wave period. By analogy with the probability of photon emission by a particle \cite{Duclous_PPCF2011} let us consider the probability of a particle losing a significant part of its energy during $T/2$ and consequently maintaining the trapped state as $Pr=1-\exp{\left(-\int_0^{T/2}\Psi dt\right)}$, where $\Psi$ is the probability of losing a great part of particle energy per unit time. During $N_{0.5}$ halves of wave periods the probability of trapping is $Pr^{N_{0.5}}$. Since $t_{10}$ is determined as the moment when the number of particles in the focus becomes an order of magnitude less than the initial number, the corresponding number of half periods is determined by the equation
\begin{equation}
	Pr^{N_{0.5}}=0.1
	\label{EqProb}
\end{equation}
and the escape time can be estimated as
\begin{equation}
	t_{10}=0.5N_{0.5}T.
	\label{Eqt10Gen}
\end{equation}
From Eq.~\ref{EqProb} it follows that $N_{0.5}=\ln{(0.1)}/\ln{\left(1-\exp{\left(-\int_0^{T/2}\Psi dt\right)}\right)}$. Substituting this expression into Eq.~\ref{Eqt10Gen} we arrive at an estimate $t_{10}\approx0.5\ln{(0.1)}T/\ln{\left(1-\exp{\left(-\int_0^{T/2}\Psi dt\right)}\right)}$. To derive rigorously the time of escape we need to compute $\int_0^{T/2}\Psi dt$ along particle trajectories. To avoid it we may assume $\int_0^{T/2}\Psi dt\approx a_1P_\mathrm{PW}^{a_2}$ and fit numerical data since usually a similar integral can be considered approximately as a power law function of wave amplitude in different field structures \cite{Fedotov_PRL2010,Bashmakov_POP2014,Mironov_PRA2021,Kostyukov_POP2016}. The best fitting of $t_{10}(P)$ (obtained numerically in the power region $P>25~PW$ where \textit{axial} ART manifests) yields $a_1\approx0.18$ and $a_2\approx0.49$, so a good approximation is
\begin{equation}
	\overline{t}_{10}(P>25~PW)=\frac{0.5\ln{(0.1)T}}{\ln{\left(1-\exp\left(-0.18P_\mathrm{PW}^{0.49}\right)\right)}}\approx\frac{-1.15T}{\ln{\left(1-\exp{\left(-2.6\times10^{-4}a^{0.98}\right)}\right)}}
	\label{Eqt10App}
\end{equation}
shown in (Fig.~\ref{FigEscape}(d)) by the dash-dotted line. This fitting is in a good agreement with the numerical results and shows quite a steep growth of escape time as a function of power or wave amplitude. Note that in the case of the e-dipole wave, at great powers ($P_{th5}<P<300~PW$) the escape time is proportional to $P_\mathrm{PW}^{0.17}$ \cite{Bashinov_QE2019}. Such difference is due to in the e-dipole wave ART not being so strong and particle escape from the trapped state even without stochasticity of photon emission.

In order to conclude Sec~\ref{Sec3_2}-\ref{Sec3_4} we summarize the results. Based on an analysis of trajectories, escape time and averaged particle distributions we revealed different power thresholds, which correspond to changes in dynamics of the particle ensemble. The revealed thresholds correspond not only to emergence of new regimes of motion, but also redistribution of particles between them. The possible regimes of motion are the following. At powers $P<P_{th1}=0.1~\mathrm{PW}$ radiation losses are negligible and \textit{ponderomotive trapping} and \textit{ponderomotive escape} are the main regimes. At $P\approx P_{th1}$ \textit{radial} ART emerges. Such a low power threshold is caused by the specific field structure in which particles can be trapped in a strong field region even without radiative effects. NRT appears at $P\approx P_{th2}=1~\mathrm{PW}$ and \textit{axial} ART becomes possible at $P\approx P_{th4}=25~\mathrm{PW}$. Since field structures are not symmetric with respect to the surface crossing the electric field antinode and parallel to the $z$ axis, the ART regimes are different in different regions. In the vicinity of the antinode closer to the central point particles can move in the \textit{radial} ART regime, experiencing radial attraction to the antinode and axial drift outwards from the central plane. On the other side of the surface particles can move in the \textit{axial} ART regime, oscillating between the node and the antinode and getting attracted to the central plane. In \textit{axial} ART an attractor is almost formed, and is only prevented by randomness of photon emission. This results in quite a steep dependence of trapping time and escape time on wave power.

Also the ART regimes increase the angle of particle escape with respect to the $z$ axis while NRT favors axial particle escape and lead to the reduction of the energy of escaping particles. When $P>P_{th4}$, the trapping regimes mentioned above become prominent in regions at a larger distance from the axis.

\subsection{Energy and angular spectra of photons\label{Sec3_5}}

\begin{figure}
	\includegraphics[width = 1\columnwidth]{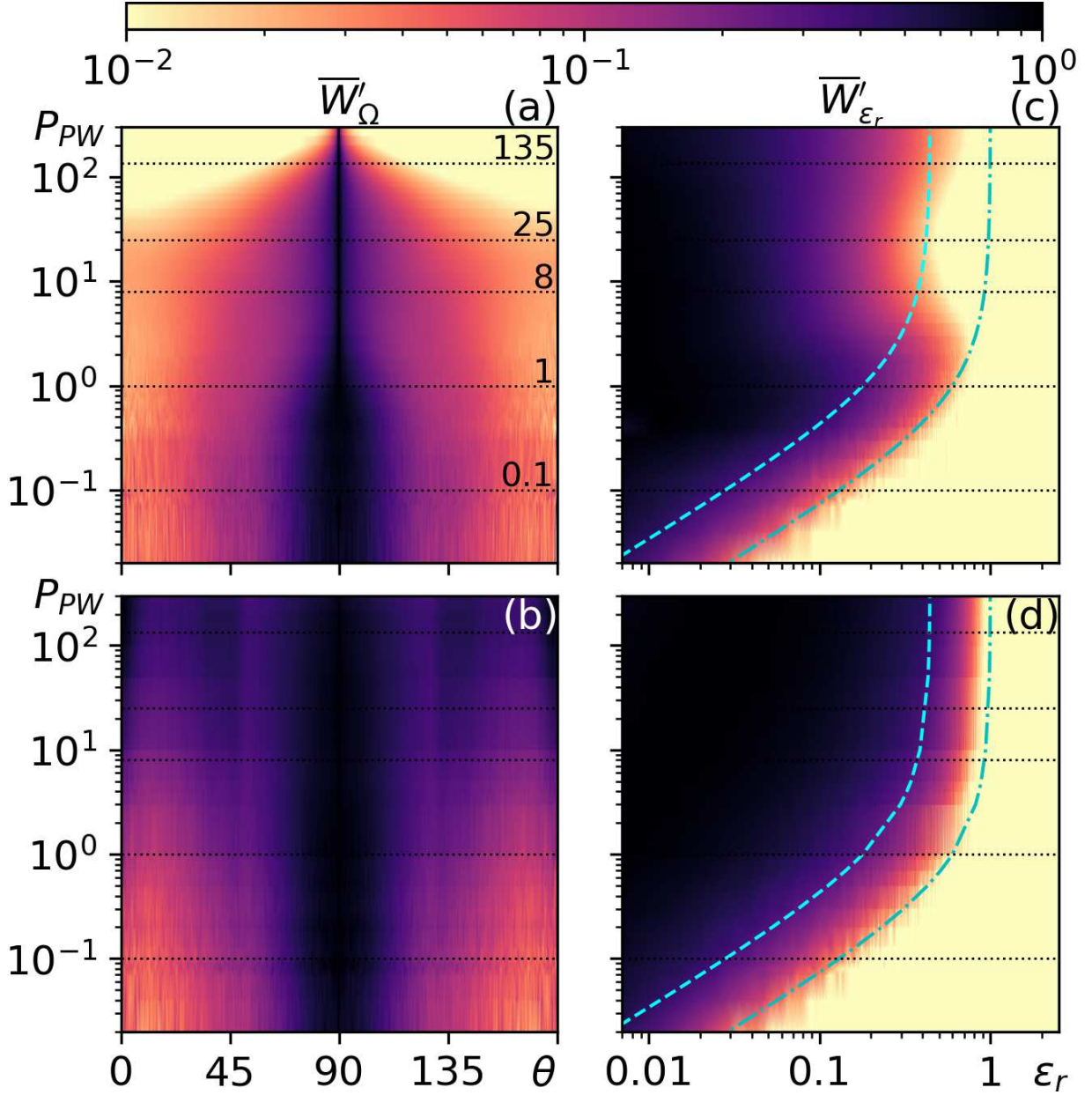}
	\caption{(Color online) Maps of (a), (b) averaged angular $\overline{W}_\Omega^{'}$ and (c), (d) averaged energy $\overline{W}_{\varepsilon_r}^{'}$ distributions of generated gamma photons escaping observation sphere. For comparison we present results obtained (a), (c) with and (b), (d) without photon recoil taken into account. Horizontal black dotted lines show previously determined power thresholds. Estimates of characteristic relative energies of photons $\varepsilon_{r,\gamma}^\mathrm{cut}$ and $\varepsilon_{r,\gamma}^\mathrm{cut2}$ emitted by particles with maximal possible and half of maximal possible energies as functions of P are shown by dash-dotted and dashed lines in (c) and (d).\label{FigMapPh}}
\end{figure}

From the practical point of view it is also important to determine the angular and energy characteristics of generated gamma photons. The significant part of total photon energy is emitted while particles oscillate in the region of strong fields, moreover photons propagate along straight lines, so they can be good characteristics of particle motion in the focus. In order to characterize escaping photons we use expressions for $\overline{W}_\Omega^{'}$ and $\overline{W}_{\varepsilon_r}^{'}$ introduced in Sec.~\ref{Sec3_3}, but we replace particle characteristics like momentum and energy with those of the photon. Opposite to the particle angular and relative energy spectra, the same characteristics for photons depend nonlinearly on power and field amplitude even without photon recoil. For example, although photon escape is mainly transverse in the whole power range with or without photon recoil (see Fig.~\ref{FigMapPh}(a),~(b)), the angular spread depends on power.

For the process of photon emission the key parameter along with particle energy is the particle’s dimensionless quantum parameter $\chi$ which determines the probability of photon emission and the photon spectrum. Though analytical derivation of the photon spectrum is challenging, it is possible to estimate the cutoff of relative photon energy $\varepsilon_{r,\gamma}^\mathrm{cut}$. The maximal possible particle energy is approximately $\varepsilon^\mathrm{cut}=a_Emc^2$, maximal transverse field is around $a$, so, according to Eqs.~\ref{EqEBamp} and \ref{EqChi} $\chi^\mathrm{cut}\approx\eta aa_E\approx1.1P_\mathrm{PW}$. The charateristic relative photon energy emitted by a particle with $\varepsilon^\mathrm{cut}$ and $\chi^\mathrm{cut}$ can be estimated according to Ref.~\cite{Berestetskii} as 
\begin{equation}
	\varepsilon_{r,\gamma}^\mathrm{cut}\approx\chi^\mathrm{cut}/(2/3+\chi^\mathrm{cut})\approx1.1P_\mathrm{PW}/(2/3+1.1P_\mathrm{PW}).
	\label{EqPhEn}
\end{equation}
For visualization we also show the characteristic relative energy of photons emitted by particles in the same fields with a twice lower energy: $\varepsilon_{r,\gamma}^\mathrm{cut2}\approx0.25\chi^\mathrm{cut}/(2/3+0.5\chi^\mathrm{cut})\approx0.27P_\mathrm{PW}/(2/3+0.55P_\mathrm{PW})$.

In the power range $P<P_{th1}$ angular and energy characteristics with and without photon recoil are very similar (Fig.~\ref{FigMapPh}) and cutoff energies are also well approximated by Eq.~\ref{EqPhEn} (Fig.~\ref{FigMapPh}(c),~(d)). Relative photon energy grows almost linearly with increasing power. Photons escape the focal region at polar angles $70^\circ<\theta<110^\circ$. Since photons with the greatest energy are emitted in the vicinity of the central plane where the main components of motion are radial and azimuthal, and axial drift is relatively slow, the maximum of averaged angular characteristics of photons is at $\theta=90^\circ$. This conclusion is relevant for the whole power range and does not depend on photon recoil.

In the power range $P_{th1}<P<P_{th2}$ changes caused by radiation losses become noticeable. When photon recoil is allowed for, angular spread begins to shrink, see Fig.~\ref{FigMapPh}(a). The situation is opposite when we do not take into account photon recoil, see Fig.~\ref{FigMapPh}(b), in this case particle trajectories do not change with increasing power, but $\varepsilon$ and $\chi$ increase. This enhances emission of photons with a greater part of particle energy at different moments including those when the angle between particle momentum and the $z$ axis is not too large. As a result, the angular spread of gamma radiation becomes wider.

This conclusion is not applicable in the case when photon recoil is allowed for. The parts of a trajectory where photon emission is more probable become more and more crucial with increasing power, because after energy loss the probability of photon emission on other parts of the trajectory decreases. Thus, the relative energy on different parts of the trajectory strongly depends on power, and the trajectory cannot maintain the same form. Particularly, in the considered power range this leads to emerging of \textit{radial} ART which causes photon recoil to become more noticeable. Photon recoil is directed approximately opposite to the particle momentum in the ultrarelativistic case \cite{Landau2}; as a result radiation losses slow down axial drift, and particles can obtain smaller axial momentum and emit photons at larger angles with respect to the $z$ axis (compare Fig.~\ref{FigMapPh}(a) and (b)). Also, although radiation losses should lead to dissipation, the number of energetic photons with relative energies $0.01<\varepsilon_r<0.1$ increases due to the new regime of motion (compare Fig.~\ref{FigMapPh}(c) and (d)).
 
At greater powers the angular spread of escaping photons decreases and becomes around $1^\circ$ at $P=300~\mathrm{PW}$ (Fig.~\ref{FigMapPh}(a)). This is mainly related to slowing down of axial drift. Note that without photon recoil the energy distribution is much wider at powers $P>5~\mathrm{PW}$.

The energy characteristic shows more variations. In the range $P_{th2}\lesssim P\lesssim P_{th4}$, while without photon recoil relative photon energies reach approximately maximum possible values, radiation losses lead to a decrease of photon energies. This is clearly visible if the edge of the relative energy distribution $\overline{W}_{\varepsilon_r}^{'}$ is compared with $\varepsilon_{r,\gamma}^\mathrm{cut}$ and $\varepsilon_{r,\gamma}^\mathrm{cut2}$ (Fig.~\ref{FigMapPh}(c) and (d)). The emerging NRT is the reason of such decrease. When \textit{axial} ART emerges at $P\gtrsim P_{th4}$, relative photon energies are increased and also reach approximately the maximum possible values  $\varepsilon_{r,\gamma}^\mathrm{cut}$.

Thus, we show that in the standing m-dipole wave generated photons propagate in transverse direction and angular spread decreases from $40^\circ$ at $P=0.01~\mathrm{PW}$ up to $1^\circ$ at $P=300~\mathrm{PW}$. Though the angular distribution does not allow clear distinguishing of thresholds of \textit{radiative trapping} regimes, at powers greater than $1~\mathrm{PW}$ due to radiation losses it becomes significantly narrower in comparison with the one obtained without photon recoil. However, if we consider the angular distribution together with the energy distribution, power thresholds of ART and NRT regimes can be determined. The dominant regime of particle motion substantially influences the cutoff value of the energy distribution. When ART regimes are strong, the cutoff energy can be around the maximal particle energy $a_Emc^2$. In the NRT regime the cutoff energy is up to 2 times lower. The average photon energy is more than an order of magnitude lower and is in the range $(0.01-0.1)a_Emc^2$ in \textit{radiative trapping} regimes.

\section{Experimental signature of radiation friction effect\label{Sec4}}

\begin{figure}
	\includegraphics[width = 1\columnwidth]{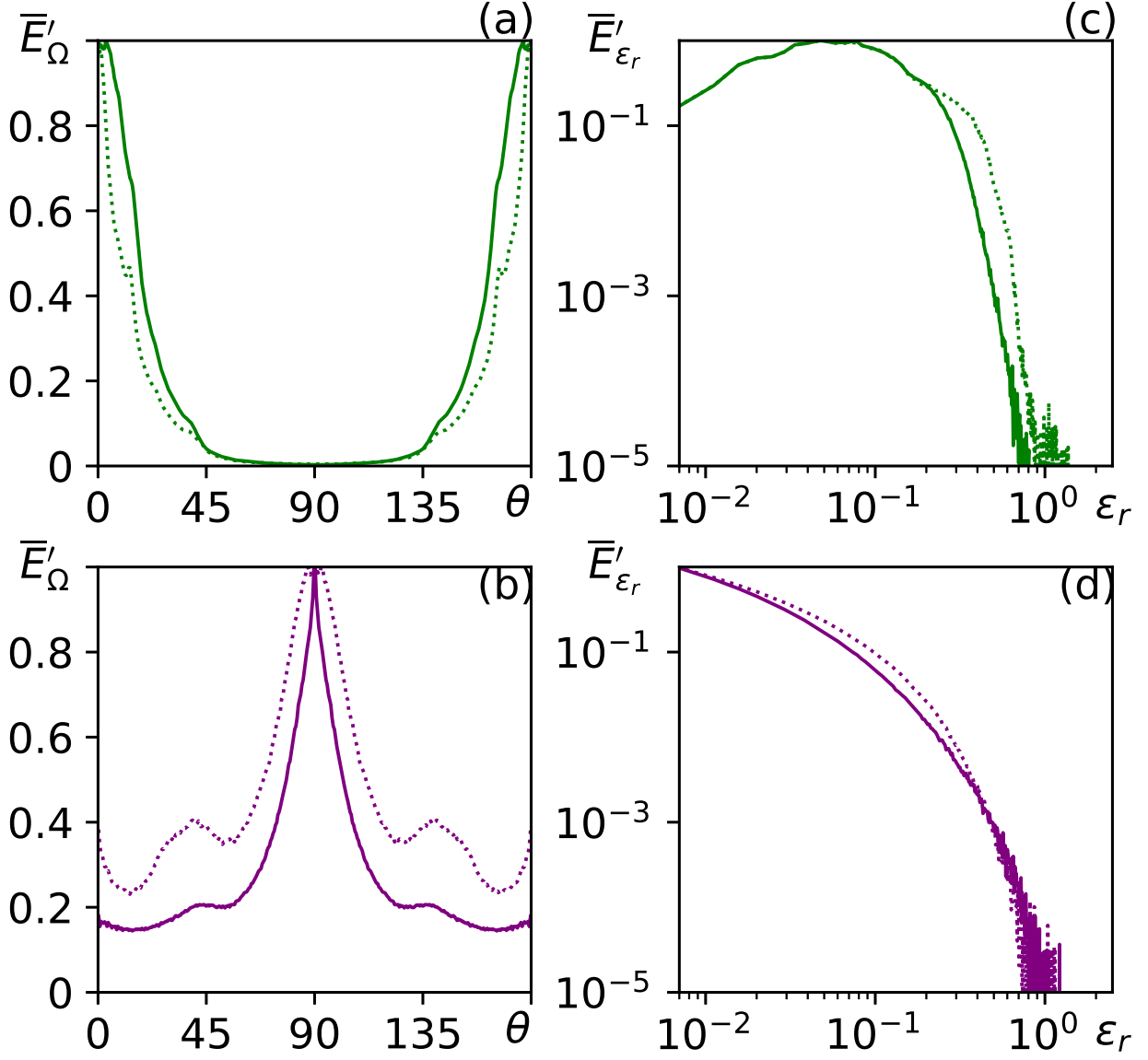}
	\caption{(Color online) (a), (b) Angular and (c), (d) energy characteristics of escaping (a), (c) electrons and (b), (d) gamma photons as a result of irradiation of cylindrical target with radius $0.25\lambda$ and density $10n_{cr}$ by the pulsed m-dipole wave with peak power $3~\mathrm{PW}$ and FWHM pulse duration of $30~\mathrm{fs}$. Solid lines correspond to simulation when photon recoil is taken into account and dotted lines are obtained without photon recoil.\label{FigTarget}}
\end{figure}

In recent years, with the advent of superpowerful lasers, there has been a lot of interest in the fundamental problem of the possibility of experimental testing of the quantum nature of the effect of the radiation friction force on the motion of relativistic particles (see for an example Refs. \cite{Cole_PRX2018,Poder_PRX2018,Wistisen_NC2018}). The main attention was paid to the case of combined use of high-power lasers and relativistic electron beams moving towards the laser wave, since in this case their interaction is the most extreme from the quantum point of view. Due to the Lorentz transformation, the frequency of scattered photons increases by a factor of $4\gamma^2$, where $\gamma$ is the relativistic gamma factor. However, results presented above show that at laser powers of only about $P_{th1}=0.1~\mathrm{PW}$, which are available in many laboratories \cite{Danson_HPLSE2019}, particle dynamics in the m-dipole wave can be strongly affected by radiative effects. The angular distribution of escaping particles and generated photons as well as their energy distributions with and without radiation losses taken into account can be distinguished. Here, we would like to propose an experiment to shed light on the quantum nature of the radiation friction effect: irradiation of a target by a converging m-dipole wave. We consider a test hydrogen-like target in the form of a nanowire with radius $0.25\lambda$ and density $n_0=10n_{cr}$, where $n_{cr}=m\omega^2/(4\pi e^2)\approx1.4\times10^{21}~\mathrm{cm^{-3}}$, irradiated by an ideal m-dipole wave with a peak power of $P=3~\mathrm{PW}$ with $\sin^2$ envelope and FWHM duration of $30~\mathrm{fs}$. In experiments the m-dipole wave can be closely mimicked by a number of linearly polarized beams \cite{Efimenko_SR2018,Gonoskov_PRL2014}.

The simulation box has sizes $6\lambda\times6\lambda\times6\lambda$ and number of cells $768\times768\times768$ along $x$, $y$ and $z$ axes, the time step is $dt=T/300$. The target is aligned along the wave symmetry axis ($z$ axis) and the initial number of particles of each type (electrons and ions) is $1.2\times10^7$. Radiation losses are modeled as random acts of photon emission with the help of the Adaptive Event Generator \cite{Gonoskov_PRE2015}. Note that the process of photon decay is also included in simulations but the rate of pair creation is slow at the considered wave power and generated pairs do not affect the laser-plasma interaction.

Initially the converging m-dipole wave pushes particles to the $z$ axis by the leading edge and compresses electrons in a wire with radius of $0.03\lambda$ and density $600n_{cr}$. After approximately $1.25T$ ions are compressed too, and an electron-ion plasma column is formed with an average radius of $0.05\lambda$. When instantaneous field amplitude in relativistic units exceeds approximately $150$, the plasma becomes transparent and the standing m-dipole wave is formed. Initially all compressed electrons drift along the z axis in the regime of \textit{ponderomotive trapping} and escape. However, when the instantaneous wave amplitude exceeds approximately $260$, some particles begin to change their regime to radial ART, deviate from the axis and become trapped for a short period of time between the first electric field node and antinode. Although ions suppress such deviation, wave fields are strong enough to overcome this suppression and allow detecting differences in particle dynamics with and without radiative effects taken into account.

In order to reveal signatures in the case of a pulsed wave we propose energy and angular characteristics very similar to those introduced in Sec.~\ref{Sec3_3} but without averaging. We take into account escaping particles and photons crossing the observation sphere with a radius of $3\lambda$ during the whole simulation time, not only after $t_{10}$ during a number of wave periods. Angular characteristic $E_\Omega^{'}$ denotes the total energy of particles or photons having crossed the observation sphere with momentum directed into an element of solid angle $d\Omega$ in the momentum space. Energy characteristic $E_{\varepsilon_r}^{'}$ denotes total energy of particles or photons having crossed the observation sphere with energy in range from $\varepsilon_r$ to $\varepsilon_r+d\varepsilon_r$. In Fig.~\ref{FigTarget} the characteristics normalized by their maximum $\overline{E}_\Omega^{'}=E_\Omega^{'}/E_\Omega^{'\mathrm{max}}$ and $\overline{E}_{\varepsilon_r}^{'}=E_{\varepsilon_r}^{'}/E_{\varepsilon_r}^{'\mathrm{max}}$ are presented. As expected from Fig.~\ref{FigMaps}(c),~(d), particles escape mainly along the axis, photons escape in the transverse direction and radiation losses lead to approximately twice the angular spread for particles and half the angular spread for photons (Fig.~\ref{FigTarget}(a),~(b)). This happens because \textit{radial} ART becomes quite strong in the petawatt range of powers. Moreover, for particles the convex shape of $\overline{E}_\Omega^{'}$ in the vicinity of $\theta=0^\circ$ and $\theta=180^\circ$ changes to a concave one due to radiation losses. In the case of photons the change is opposite. Energy characteristics $\overline{E}_{\varepsilon_r}^{'}$ of particles and photons demonstrate, as predicted (Fig.\ref{FigMaps}(e),~(f)), that radiation losses slightly decrease the number of particles and photons with high energies (Fig.~\ref{FigTarget}(c),~(d)). The reason for this is the emerging NRT regime which favors cooling of particles.

Thus, we show that with petawatt lasers there is a possibility to experimentally detect signatures of quantum radiative effects based on angular and energy spectra of particles and photons. However, it should be noted that in experiments with higher powers observation of radiative effects in particle dynamics can be obstructed by gamma photon decays into electron-positron pairs which can strongly redistribute particles within focus. As a result, the spectra of particles and photons can be changed significantly mainly due to pair creation. This can happen at laser powers of about $10~\mathrm{PW}$ and up \cite{Bashinov_ArX2021}.

\section{Conclusion\label{Sec5}}

We have considered particle motion in the standing m-dipole wave in a wide range of powers from $0.01~\mathrm{PW}$ to $300~\mathrm{PW}$ and showed that the motion is distinguished by a number of unique properties. First, particle motion in the m-mode fields is not planar as in the case of the e-dipole wave. Second, particles experience a centrifugal force due to the azimuthal component of the electric field, and this force plays an important role in the particle trapping effects. Third, \textit{ponderomotive trapping}, which is distinctive for the m-dipole wave in the central region, impedes escape of particles from this region, providing the opportunity for them to accumulate influence of radiative effects and reducing the threshold of the radiation-dominated regime.

In order to determine possible modes of particle motion qualitatively and quantitatively, we have proposed averaged characteristics based on the time evolution of particle ensemble, its spatial distribution, as well as energy and angular spectra. Based on these characteristics we have determined that even at subpetawatt powers radiative effects become noticeable owing to \textit{ponderomotive trapping} in the region of the strongest field. At such powers radiation losses lead to \textit{radial anomalous radiative trapping} when in the radial direction particles are attracted to the electric field antinode and in the axial direction they drift outwards from the focus. At powers greater than $1~\mathrm{PW}$ \textit{normal radiative trapping} emerges, leading to trapping and cooling of accelerated particles in the vicinity of the second electric field node. If power exceeds $25~\mathrm{PW}$, \textit{axial anomalous radiative trapping} becomes possible between the electric field antinode and its second node. Since field structures are asymmetric with respect to the electric field antinode, the properties of motion in \textit{axial} and \textit{radial} ART regimes is different. In the \textit{axial} ART regime motion is limited in the radial direction and is characterized by axial attraction to the focus. This is the unique regime in highly inhomogeneous fields that could result in formation of an attractor, which is only broken by randomness of photon emission. Nevertheless, trapping time is quite a steep function of power, so remarkable long-term trapping is possible at powers above $25~\mathrm{PW}$. Moreover, at such powers the \textit{radiative trapping regimes} become apparent at a larger distance from the focus, which due to retrapping also increases escape time.

We have shown that particles escape the focus mainly in the axial direction, however, ART regimes change axial escape to transverse escape and can significantly suppress the first one at powers of about $100~\mathrm{PW}$. In turn, NRT favors axial escape and causes near saturation of the average energy of escaping particles at the level of $100~\mathrm{MeV}$ at powers above $1~\mathrm{PW}$. We have also determined that particles emit gamma photons mainly perpendicular to the symmetry axis and radiation losses narrow the photon distribution over the polar angle from $30^\circ$ at $0.1~\mathrm{PW}$ to $1^\circ$ at $300~\mathrm{PW}$. Energy distribution of photons is more sensitive to regimes of particle motion. At subpetawatt powers \textit{radial} ART increases the average photon energy, while the ratio of the maximal photon energy to the maximal particle energy is not influenced by radiation losses and increases with power up to about 0.8. In the power range above $1~\mathrm{PW}$ NRT becomes dominant and decreases this ratio to about 0.4 at $25~\mathrm{PW}$. At greater powers \textit{axial} ART increases this ratio, which reaches about 0.7 at above $100~\mathrm{PW}$.

Based on the obtained results we have proposed a schematic of a laboratory experiment on irradiation of a nanowire by a converging m-dipole wave. A total power of about $1~\mathrm{PW}$ and pulse duration of $30~\mathrm{fs}$ has been predicted to be sufficient in order to detect signatures of radiative effects.

\section*{Acknowledgments}

This work was funded by the Ministry of Education and Science of the Russian Federation under contract No. 075-15-2021-633.

\bibliography{Bibl}

\end{document}